\documentclass[10pt,journal,compsoc]{IEEEtran}
\usepackage{amsmath}
\usepackage{amssymb}
\usepackage{booktabs}
\usepackage{epsfig}
\usepackage{enumitem}
\usepackage{multirow}
\usepackage{setspace}
\usepackage{bigstrut}
\usepackage{microtype}      % microtypograph
\usepackage[american]{babel}
\usepackage[caption=false]{subfig}
\usepackage{url}
\usepackage{bm}
\usepackage{color}
\newcommand{\eg}{\emph{e.g.}}

\newcommand{\ie}{\emph{i.e.}}
\newcommand{\etal}{\emph{et al.}}

\graphicspath{{figure/}}

\ifCLASSOPTIONcompsoc
  \usepackage[nocompress]{cite}
\else
  \usepackage{cite}
\fi

\ifCLASSINFOpdf
\else
\fi

\begin{document}

%\title{Large-Scale Fine-Grained Image Retrieval with Attribute-Aware Hash Codes}
\title{Attribute-Aware Deep Hashing with Self-Consistency for Large-Scale Fine-Grained Image Retrieval}

\author{
	Xiu-Shen Wei,~\IEEEmembership{Member,~IEEE}, Yang Shen, Xuhao Sun, Peng Wang,~\IEEEmembership{Member,~IEEE}, Yuxin Peng,~\IEEEmembership{Senior Member,~IEEE}
\IEEEcompsocitemizethanks{\IEEEcompsocthanksitem {X.-S. Wei is with School of Computer Science and Engineering, and Key Laboratory of New Generation Artificial Intelligence Technology and Its Interdisciplinary Applications, Southeast University, China. Y. Shen and X. Sun are with School of Computer Science and Engineering, Nanjing University of Science and Technology, China. P. Wang is with School of Computer Science and Engineering, University of Electronic Science and Technology of China, China. Y. Peng is with Wangxuan Institute of Computer Technology, and National Key Laboratory for Multimedia Information Processing, Peking University, China. Y. Peng is the corresponding author.}
\IEEEcompsocthanksitem{This work was supported by National Key R\&D Program of China (2021YFA1001100), National Natural Science Foundation of China under Grants (62272231, 62132001, 61925201), Natural Science Foundation of Jiangsu Province of China under Grant (BK20210340), the Fundamental Research Funds for the Central Universities (No. NJ2022028), and CAAI-Huawei MindSpore Open Fund.}}
%~\IEEEmembership{Member,~IEEE,}
}

\markboth{ACCEPTED BY IEEE TPAMI}%
{Wei \MakeLowercase{\textit{et al.}}: \textsc{A$^2$-Net}: Learning Attribute-Aware Hash Codes for Large-Scale Fine-Grained Image Retrieval}

\IEEEtitleabstractindextext{%
\begin{abstract}
Our work focuses on tackling large-scale fine-grained image retrieval as ranking the images depicting the concept of interests (\ie, the same sub-category labels) highest based on the fine-grained details in the query. It is desirable to alleviate the challenges of both fine-grained nature of small inter-class variations with large intra-class variations and explosive growth of fine-grained data for such a practical task. In this paper, we propose attribute-aware hashing networks with self-consistency for generating attribute-aware hash codes to not only make the retrieval process efficient, but also establish explicit correspondences between hash codes and visual attributes. Specifically, based on the captured visual representations by attention, we develop an encoder-decoder structure network of a reconstruction task to unsupervisedly distill high-level attribute-specific vectors from the appearance-specific visual representations without attribute annotations. Our models are also equipped with a feature decorrelation constraint upon these attribute vectors to strengthen their representative abilities. Then, driven by preserving original entities' similarity, the required hash codes can be generated from these attribute-specific vectors and thus become attribute-aware. Furthermore, to combat simplicity bias in deep hashing, we consider the model design from the perspective of the self-consistency principle and propose to further enhance models' self-consistency by equipping an additional image reconstruction path. Comprehensive quantitative experiments under diverse empirical settings on six fine-grained retrieval datasets and two generic retrieval datasets show the superiority of our models over competing methods. Moreover, qualitative results demonstrate that not only the obtained hash codes can strongly correspond to certain kinds of crucial properties of fine-grained objects, but also our self-consistency designs can effectively overcome simplicity bias in fine-grained hashing. 
\end{abstract}

}

\maketitle

%\tableofcontents

\IEEEdisplaynontitleabstractindextext

\IEEEpeerreviewmaketitle

\section{Introduction}

\IEEEPARstart{F}{ine}-grained image retrieval~\cite{XSsurveyPAMI} in computer vision and pattern recognition aims to retrieve images belonging to multiple subordinate categories of a super-category (\emph{aka} a meta-category), \eg, different species of animals/plants~\cite{inat2017}, different models of cars~\cite{cars}, different kinds of retail products~\cite{rpc}, etc. Its key challenge therefore lies in understanding fine-grained visual differences that subtly distinguish objects that are highly similar in overall appearance, but differ in \emph{fine-grained} features. Also, fine-grained retrieval still demands ranking all the instances so that images depicting the same sub-category label are ranked highest based on the fine-grained details in the query.

\begin{figure*}[t!]
\centering
{\includegraphics[width=0.95\textwidth]{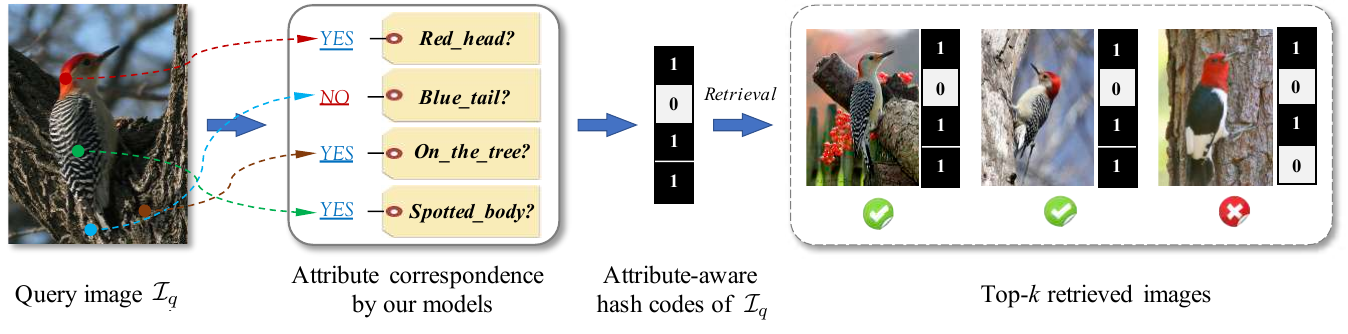}}
\vspace{-0.5em}
\caption{Key idea of our models, as well as the main process of fine-grained hashing based on our \emph{attribute-aware} hash codes. Concretely, regarding a query image $\mathcal{I}_q$ of \texttt{Red bellied Woodpecker}, after returning all the correct results, a fine-grained image belonging to \texttt{Red headed Woodpecker} closest to the query image in terms of Hamming distance is also retrieved.}
\label{fig:idea}
\end{figure*}

In particular, with the explosive growth of fine-grained data in real applications~\cite{Birdsnap14,deepfashion16,vegfru,inat2017,rpc}, fine-grained hashing, as a promising solution for dealing with large-scale fine-grained retrieval tasks, has proven to be able to greatly reduce the storage cost and increase the query speed~\cite{exchnet,FGhashTIP} benefiting from the learned compact binary hash code representations. While previous work, \eg,~\cite{exchnet,FGhashTIP}, achieved good retrieval performance, however, the hash codes they generated lack semantic correspondence, \ie, they did not capture \emph{fine-grained attributes}. These attributes, such as head colors, tail colors, male, female, living habits, are as effective descriptions of fine-grained objects, that can be understood by both humans and computers. In this paper, to establish an explicit correspondence between hash codes and visual attributes to simultaneously improve large-scale fine-grained retrieval accuracy and integrate interpretation into deep learning based hash methods, we propose unified Attribute-Aware hashing Networks, termed as \textsc{A$^2$-Net} and \textsc{A$^2$-Net$^{++}$}, for achieving these goals (cf. Figure~\ref{fig:idea}).

In both \textsc{A$^2$-Net} and \textsc{A$^2$-Net$^{++}$}, considering huge labor cost of supervised attribute annotations, we adopt an unsupervised setting to automatically capture discriminative visual attributes from still images and then associate the learned hash code representations to these attributes. Therefore, a hash bit of learned hash codes could be both discriminative and interpretable. Additionally, thanks to the unsupervised setting, the attributes derived from our models will be not restricted to pre-defined attributes like supervised-based attribute learning methods~\cite{dualawarericcv15,facepami3310,audioaaai2018,awapami}. Moreover, it can distill the most useful properties of fine-grained objects as attribute-aware hash codes in such an end-to-end trainable manner for accurate retrieval among multiple similar subordinate categories.

More specifically, as the overall framework shown in Figure~\ref{fig:framework}, our \textsc{A$^2$-Net}, as the base model w.r.t. \textsc{A$^2$-Net$^{++}$}, consists of a fine-grained representation learning module and an attribute-aware hash codes generation module. It first leverages attention mechanisms to model fine-grained patterns in terms of both global-level deep features $\bm{T}_i$ and local-level cues $\bm{T}_i^c$ from input image $\mathcal{I}_i$. Then, the appearance-specific features of these visual patterns $\bm{T}$ are aggregated and translated into semantic-specific representations $\bm{x}_i$. After that, we formulate the aforementioned unsupervised attribute learning as a reconstruction task of projecting $\bm{x}_i$ to an attribute vector $\bm{v}_i$ by performing an encoder-decoder structure network. Therefore, it can be expected that in the high-level attribute space, $\bm{v}_i$ could correspond to certain kinds of nameable properties of fine-grained objects. Moreover, a feature decorrelation constraint is further introduced upon $\bm{v}_i$ to both enhance the discriminative ability and remove the redundancy among these dimensions of attribute-specific features. Finally, our attribute-aware hash codes $\bm{u}_i$ are generated from $\bm{v}_i$ by conducting the hash code learning procedure.

It is noticed that the most crucial mechanism to generate the attribute-aware internal latent representations of inputs is the encoder-decoder structure in our proposal. It realizes a reconstruction paradigm by reconstructing the holistic feature representation $\bm{x}_i$. Beyond acquiring the implicit representations, we also find that such a structure explicitly conforms to the principle of self-consistency~\cite{Yima2022} in deep networks. The self-consistency principle encourages an intelligent system to seek a most self-consistent model for observations of the external world by minimizing the internal discrepancy between the observed and the regenerated, which enables the system to learn as comprehensive information as possible that is sensed. Therefore, to verify the merits of our proposal, we conduct preliminary experiments on a constructed dataset, \ie, \emph{MNIST-CIFAR}~\cite{simpBiasNIPs20}. This special dataset consists of images by vertically concatenating \emph{MNIST}~\cite{mnist} images onto \emph{CIFAR-10}~\cite{cifar} images, with ten classes where each class is a combination of a unique \emph{MNIST} class and \emph{CIFAR-10} class. After networks training converges on this dataset, we can visualize the learned activations on the MNIST-CIFAR images to show whether there is more attention on complex patterns comparing to the \emph{CIFAR} parts or more attention on simple patterns comparing to the \emph{MNIST} parts. As expected, it is clear that, our proposal, which exhibits self-consistency, attends to more complex and comprehensive patterns than the vanilla baseline. More importantly, these observations validate the simplicity bias phenomenon~\cite{simpBiasNIPs20} of networks in deep hash learning for the first time to our best knowledge. Detailed preliminary findings and discussions can refer to Section~\ref{sec:enhanceSC}.

Therefore, to overcome the simplicity bias and then improve the retrieval accuracy, we propose to further enhance the self-consistency in our models. Specifically, we realize this idea by equipping an additional simple but effective path upon the basic \textsc{A$^2$-Net}, where the original input image $\mathcal{I}_i$ is required to be reconstructed. Such a model with both reconstruction features and images as its necessary functions can more faithfully reflect self-consistency, thereby alleviating simplicity bias and achieving better hash retrieval accuracy, which is thus termed as the advanced \textsc{A$^2$-Net$^{++}$} model.

To evaluate our models, we conduct extensive experiments on five fine-grained retrieval benchmark datasets, \ie, \textit{CUB200-2011}~\cite{WahCUB200_2011}, \textit{Aircraft}~\cite{airplanes}, \textit{Food101}~\cite{food101}, \textit{NABirds}~\cite{nabirds15} and \textit{VegFru}~\cite{vegfru}, for validating the effectiveness and on two generic image datasets, \ie, \emph{NUS-WIDE}~\cite{NUSWIDE} and \emph{COCO}~\cite{cocoeccvMS}, for investigating the universality. Additionally, two challenging empirical settings, \ie, zero-shot fine-grained retrieval on three aforementioned fine-grained datasets  and cross-domain fine-grained retrieval on the \emph{Clothing} dataset~\cite{dualawarericcv15}, are also incorporated in comparisons for proving the generalization ability, as well as the models' robustness. Quantitative results of retrieval accuracy on these datasets and settings show that our proposed models obviously and consistently outperform existing state-of-the-art methods. Also, the ablation studies of these crucial components in our models validate their own effectiveness. In particular, to evaluate the attribute-aware retrieval performance, which is regarded as one of the most important characteristics of our models, we perform both qualitative analyses and quantitative attribute matching to demonstrate that our learned hash bits have strong correspondences to visual attributes of fine-grained objects, even without employing attribute supervisions or part-level annotations. In addition, more visualization results of simplicity bias on these fine-grained retrieval datasets are presented, which demonstrate that our models are able to effectively alleviate simplicity bias for fine-grained hashing.

Note that a preliminary version of this paper was published as a conference paper (Spotlight Presentation)~\cite{XSA2Net} in \emph{Advances in Neural Information Processing Systems 2021}. In this journal paper, we have made significant extensions. First, we systematically investigate the simplicity bias~\cite{simpBiasNIPs20} problem in deep hashing, and provide in-depth discussions from the self-consistency principle~\cite{Yima2022} perspective. Second, we propose an advanced model, \ie, \textsc{A$^2$-Net$^{++}$}, which not only achieves much better fine-grained hashing retrieval accuracy but also confirms the conjecture about the advantage of the self-consistency design in our models. Third, we conduct more comprehensive experiments and analyses to evaluate our models from many diverse aspects, such as effectiveness, robustness, generalization ability, interpretation, etc.

The rest of the paper is organized as follows. Section~\ref{sec:related} retrospects the related work. Section~\ref{sec:approach} introduces the details of our proposed models, as well as the investigation about simplicity bias in deep hashing. Experiments and analyses of both qualitative and quantitative aspects are provided in Section~\ref{sec:experiments}. Section~\ref{sec:conc} presents conclusions and future work.

\section{Related Work}\label{sec:related}

We briefly review the related work in the following four aspects, \ie, fine-grained image retrieval, learning to hash, the studies about visual attributes in computer vision problems, and simplicity bias in deep neural networks.

\subsection{Fine-Grained Image Retrieval}

Fine-grained image retrieval as an integral part of fine-grained image analysis~\cite{XSsurveyPAMI} has gained more and more traction in recent years~\cite{XSA2Net,exchnet,Wei16scda,xiawuijcai18,xiawuaaai19,SBIRbmvc14,thatshoecvpr,jigsawCVPR20}. What makes it challenging is that objects of fine-grained images have only subtle differences, and often largely vary in pose, scale, and orientation or can exhibit cross-modal differences (\eg, sketch-based retrieval~\cite{SBIRbmvc14}).

Depending on the type of query image, the most studied areas of fine-grained image retrieval can be separated into two groups: fine-grained content-based image retrieval (FG-CBIR) and fine-grained sketch-based image retrieval (FG-SBIR). More specifically, in FG-CBIR, unsupervised learning based~\cite{Wei16scda} and supervised learning based methods~\cite{xiawuijcai18,xiawuaaai19,pce2020imavis} were developed from different perspectives for handling fine-grained retrieval tasks, \eg, localizing fine-grained parts~\cite{Wei16scda}, enhancing intra-class separability with inter-class compactness~\cite{xiawuaaai19}, and reducing the confidence of the fine-grained predictions~\cite{pce2020imavis}, etc. While, FG-SBIR needs to not only capture fine-grained characteristics present in the sketches, but also possess the ability to traverse the sketch and image domain gap. In the literature of FG-SBIR, the earlier work, \eg,~\cite{thatshoecvpr,zhang2018generative,TIPFGSBIR17}, were mostly based on Siamese-triplet networks~\cite{siamesenips93} to tackle the aforementioned challenges. Recently, some work tried to incorporate the advances of recent progress in self-supervised learning~\cite{simclr} and attention mechanisms~\cite{natureneuro02} for further improving the retrieval accuracy of FG-SBIR, \eg,~\cite{jigsawCVPR20,sain2020cross}.

However, although these fine-grained retrieval methods achieved good results, they still have the limitations in the face of large-scale data, \ie, the searching time for the exact nearest neighbor is typically expensive or even impossible for the given queries. To alleviate this issue, fine-grained hashing, which aims to generate compact binary codes to represent fine-grained images, as a promising direction has attracted the attention in the fine-grained community very recently~\cite{exchnet,jin2020deep,tipxinshunTIP}. More specifically, ExchNet~\cite{exchnet} was the first to define the fine-grained hashing task and develop a fine-grained tailored method to firstly locate discriminative object parts and further learn binary hash codes for representing fine-grained images. In the same period, DSaH~\cite{jin2020deep} was proposed to automatically mine salient regions and learn semantic-preserving hash codes simultaneously. Recently, FISH~\cite{tipxinshunTIP} was developed with a double-filtering mechanism and a proxy-based loss function for handling fine-grained hashing. Unfortunately, the learned hash bits of these methods lack any semantics which are more meaningful to fine-grained objects, and thus lack the model interpretability. Compared with them, our models can not only outperform previous fine-grained hashing methods, but more importantly, the learned hash codes of our models are \emph{attribute-aware}, \ie, our hash bits have strong correspondence to semantic visual properties that are useful for fine-grained image retrieval.

\begin{figure*}[t!]
\centering
%\vspace{0.12em}
	{\includegraphics[width=0.95\textwidth]{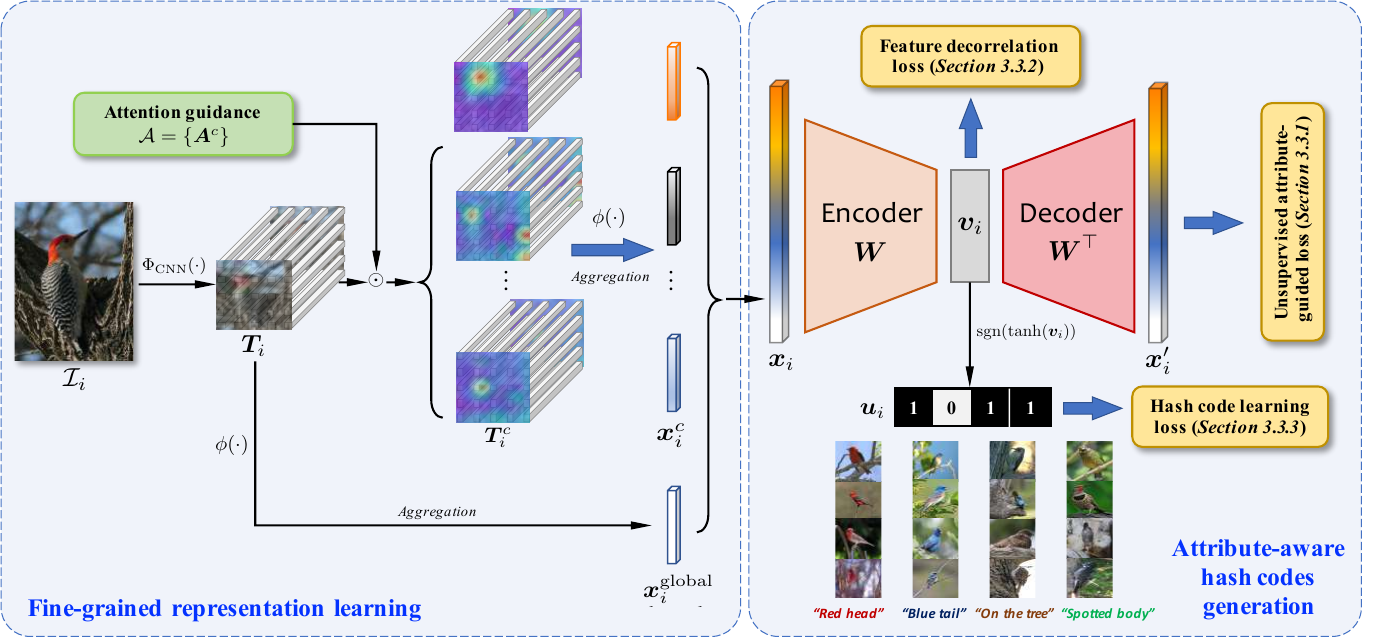}}
\caption{Overall framework of the basic \textsc{A$^2$-Net} model, which consists of two crucial modules, \ie, fine-grained representation learning and attribute-aware hash codes generation. The whole network can be end-to-end trainable, and is generally driven by the unsupervised attribute-guided reconstruction loss, the feature decorrelation loss and the hash code learning loss, cf. Section~\ref{sec:a2codes}.}
\label{fig:framework}
\end{figure*}

\subsection{Learning to Hash}

Hashing~\cite{hashsurvey} is a widely-studied solution to approximate nearest neighbor search, which transforms the data item to a short code consisting of a sequence of bits (\ie, hash codes). The research efforts of hashing can be categorized into two groups, including data-independent hashing (\emph{aka} locality sensitive hashing~\cite{fastkdd17th,vldb2007,densifyingicml14}) and data-dependent hashing (\emph{aka} learning to hash~\cite{csqCVPR20,onelossforall,qingyuanAAAI18,shen2015supervised,mutualhashtpami,cvpr18kunhehash}).

Specifically, locality sensitive hashing methods attempted to adjust hash learning from different perspectives, \eg, the theory or machine learning views, to name a few: proposing random hash functions satisfying local sensitive property~\cite{fastkdd17th}, developing better search schemes~\cite{vldb2007}, providing faster computation of hash functions~\cite{densifyingicml14}, etc. While, compared with locality sensitive hashing methods, since data-dependent hashing methods learn hash functions from a specific dataset to achieve similarity preserving, they can generally obtain superior retrieval accuracy. Especially for capitalizing on advances in deep learning, many well-performing methods were proposed to integrate feature learning and hash code learning into an end-to-end framework based on deep networks, \eg,~\cite{csqCVPR20,onelossforall,qingyuanAAAI18,mutualhashtpami,cvpr18kunhehash}. In particular, very recently researchers in the vision community have begun to pay attention to the more challenging and practical hashing task, \ie, fine-grained hashing~\cite{exchnet,jin2020deep,tipxinshunTIP}. Distant from these previous work, our proposed models equip these learned hash codes with strong correspondence to visual attributes, and meanwhile are able to combat simplicity bias~\cite{simpBiasNIPs20} with our self-consistency designs for dealing with large-scale fine-grained image retrieval.

\subsection{Visual Attributes}

Attributes are typically mid-level semantic properties of objects~\cite{learnVAnips}, such as colors (\eg, ``red'', ``blue''), texture (\eg, ``striped'', ``spotted''), or even life habits of animals (\eg, ``living on the tree'', ``living in the water''). Visual attributes have exhibited their impact for strengthening various vision tasks, including facial verification~\cite{facepami3310}, fine-grained categorization~\cite{audioaaai2018}, zero-shot transfer~\cite{awapami}, scene understanding~\cite{SUNIJCV}, and so on. Most of the previous attribute learning methods are supervised by costly human-generated annotations and also are dependent on pre-defined attribute labels, \eg,~\cite{dualawarericcv15,facepami3310,audioaaai2018,awapami}. In consequence, for large-scale problems, these supervised methods might be not feasible due to the restriction caused by the cumbersomely obtained attribute annotations. Moreover, even for some tasks, their visual  attributes are quite hard to define. In this paper, to alleviate the aforementioned issues, we propose the \textsc{A$^2$-Net} and \textsc{A$^2$-Net$^{++}$} models to formulate an \emph{unsupervised learning} structure to project the learned visual features into an attribute space where it finally generates attribute-aware binary hash codes. Compared with previous attribute learning methods, our models are independent with pre-defined attribute labels, and could automatically learn discriminative attribute-aware hash codes in a unified end-to-end trainable fashion. Furthermore, our models can not only correspond hash bits to visual attributes tailored for fine-grained objects, which shows significant improvements of retrieval accuracy, but also offer an intuitive way of deep hashing interpretation.

\subsection{Simplicity Bias in Neural Networks}

Simplicity bias (SB) refers to the inherent tendency of human perception and cognition to favor simple and easily interpretable explanations or representations over complex ones. In the literature of machine learning, SB, \ie, the tendency of supervised neural networks to find simple patterns, was proposed to analyze the generalization of neural networks~\cite{simpBiasNIPs20,huh2021low}. In particular, in~\cite{simpBiasNIPs20}, it originally showed that neural networks trained with stochastic gradient descent are biased to learning the simplest predictive features in the data, while overlooking the complex but equally-predictive patterns. Inspired by this, Teney~\etal~\cite{OODcvpr22} demonstrated that SB can be mitigated through a diversity constraint and explored how to improve out of distribution generalization by overcoming SB.

In the context of fine-grained image retrieval, SB can play a significant role in shaping our perception and interpretation of detailed visual information. Fine-grained image retrieval deals with the retrieval tasks of objects or entities within a specific category that exhibit subtle and intricate visual differences. These differences may be challenging to discern and require careful examination of fine details. However, SB leads us to overlook or underestimate the importance of these fine-grained details. Our cognitive inclination towards simplicity may push us towards making more general and coarse classifications, focusing on prominent features or easily distinguishable patterns while neglecting the nuances and subtle variations that are essential for accurate fine-grained retrieval. In this paper, to our best knowledge, we are the first to investigate the SB problem in the fine-grained image retrieval task, and propose methods from the model's self-consistency perspective~\cite{Yima2022} to effectively mitigate SB in fine-grained image-oriented hash code learning.

\section{Methodology}\label{sec:approach}

In this section, we introduce the overall framework, notations of the proposed models, as well as elaborating its crucial modules and the enhanced self-consistency designs for further alleviating the simplicity bias.

\subsection{Overall Framework and Notations}

As illustrated in Figure~\ref{fig:framework}, our basic \textsc{A$^2$-Net} model consists of two crucial modules, \ie, a fine-grained representation learning module and an attribute-aware hash codes generation module. Given an input image $\mathcal{I}_i$, based on its corresponding deep activation tensor $\bm{T}_{i}\in \mathbb{R}^{C\times H\times W}$ extracted by a backbone CNN, a set of attention guidance $\mathcal{A}=\{\bm{A}^c\}$ is learned for capturing fine-grained tailored local patterns $\bm{T}_{i}^c$ from $\bm{T}_{i}$. To distill semantical cues and further generate the final attribute-aware binary hash codes, we propose to transform these appearance-specific features $\bm{T}$ towards semantic-specific representations $\hat{\bm{T}}$ by performing a transform network $\phi(\cdot)$. After aggregating $\hat{\bm{T}}$, the obtained attentive local-level features $\bm{x}^c_i$ are associated with the global-level feature $\bm{x}_i^{\rm global}$ to form as a holistic feature representation $\bm{x}_i$. In order to generate attribute-aware binary hash codes, we conduct a reconstruction paradigm to project $\bm{x}_i$ as $\bm{v}_i$ in an attribute space where its data point corresponds to an attribute vector w.r.t. a certain kind of nameable properties of fine-grained objects (\eg, ``red head'' or ``spotted body''). Furthermore, with the aid of feature decorrelation, $\bm{v}_i$ is expected to be more discriminative by removing redundant correlation information. Finally, hash code learning is performed upon $\bm{v}_i$ to obtain the final attribute-aware binary codes $\bm{u}_i$.

\subsection{Fine-Grained Representation Learning}\label{sec:FGRL}

Attention plays an important role in human perception~\cite{pami98attention,natureneuro02}, and humans exploit a sequence of partial glimpses and selectively focus on salient parts of an object or a scene in order to better capture visual structure~\cite{nipshiton2010}. Inspired by this, we incorporate the attention mechanism into representation learning to capture fine-grained local patterns for distinguishing subtle differences between these subordinate categories.

Concretely, we extract the deep feature of its input image $\mathcal{I}_i$ via a backbone CNN model $\Phi_{\rm CNN}(\cdot)$ by
\begin{equation}
\bm{T}_i = \Phi_{\rm CNN}(\mathcal{I}_i)\in \mathbb{R}^{C\times H\times W}\,.
\end{equation}
Then, based on $\bm{T}_i$, multiple attention guidances ($\bm{A}^c\in \mathbb{R}^{H\times W}$) are generated as a set of attention maps, \ie, $\mathcal{A}$. The attention guidance $\bm{A}^c$ is designed to evaluate which deep descriptors~\cite{Wei16scda} in these $H\times W$ cells should be attended or even overlooked by conducting
\begin{equation}
\bm{T}_i^c = \bm{A}^c \odot \bm{T}_i\,,
\end{equation}
where $\odot$ is the element-wise Hadamard product. To obtain the final attribute-aware binary codes, it is desirable to transform these appearance-specific (\ie, low-level) features $\bm{T}$ to semantic-specific (\ie, mid-level) representations which are closer to the attribute space. Thus, a transforming network $\phi(\cdot)$, which is equipped with a stack of convolution layers, is performed on $\bm{T}$ as follows:
\begin{eqnarray}
\hat{\bm{T}}_i^c = \phi(\bm{T}_i^c ; \theta_{\rm local})\,,\\
\hat{\bm{T}}_i = \phi(\bm{T}_i; \theta_{\rm global})\,,
\end{eqnarray}
where $\theta$ presents the parameters of the corresponding transforming networks w.r.t. $\bm{T}_i^c$ and $\bm{T}_i$, respectively. Then, we aggregate $\hat{\bm{T}}_i^c$ and $\hat{\bm{T}}_i$ by conducting global average-pooling and correspondingly obtain the attentive local-level features $\bm{x}_i^c$, as well as the global-level feature $\bm{x}_i^{\rm global}$. The holistic feature representation w.r.t. the input image $\mathcal{I}_i$ is achieved by concatenating both $\bm{x}_i^c$ and $\bm{x}_i^{\rm global}$, \ie, $\bm{x}_i = \left[\bm{x}_i^c; \bm{x}_i^{\rm global}\right] = F(\mathcal{I}_i; \Theta) \in \mathbb{R}^d$. Note that, we hereby abstract the aforementioned fine-grained feature learning process as a function $F(\mathcal{I}_i; \Theta)$ associated with parameters $\Theta$.

\subsection{Attribute-Aware Hash Codes Generation}\label{sec:a2codes}

How to generate attribute-aware hash codes is the key of our \textsc{A$^2$-Net} model. We elaborate it in the following three aspects, \ie, unsupervised attribute-guided learning, attribute-specific feature decorrelation, and hash code learning.

\subsubsection{Unsupervised Attribute-Guided Learning}\label{sec:UAGL}

In real-applications, especially for the large-scale and fine-grained tasks, attribute annotations are always infeasible, which limits the learning process to be conducted in an unsupervised setting. While, in the literature, the main goal of unsupervised learning is to capture regularities in data for the purpose of extracting useful representations or for restoring corrupted data~\cite{lecunAE07}. Many unsupervised methods explicitly produce \emph{internal latent units or codes}, from which the data is to be reconstructed.

Inspired by this, we develop an unsupervised attribute-guided reconstruction component to project the holistic representation $\bm{x}_i$ of $\mathcal{I}_i$ into a latent space, \ie, the attribute space $\mathcal{V}$. In $\mathcal{V}$, its high-level vectors are designed to have certain desirable properties, \eg, corresponding to semantic properties of fine-grained objects (\emph{aka} ``fine-grained attributes'').

More specifically, in our \textsc{A$^2$-Net}, the unsupervised attribute-guided learning is realized by a reconstruction paradigm with an encoder-decoder structure, as shown in Figure~\ref{fig:framework}. Concretely, given a batch of $n$ training data $\mathcal{I}_i$, their holistic representations $\bm{X}=\{\bm{x}_1;\bm{x}_2; \ldots;\bm{x}_n\}\in \mathbb{R}^{d\times n}$ can be obtained as aforementioned. By formulation, the encoder projects $\bm{X}$ into the attribute space $\mathcal{V}$ with a projection matrix $\bm{W}\in \mathbb{R}^{k\times d}$ to get an internal latent representation $\bm{V}\in \mathbb{R}^{k\times n}$ w.r.t. $\bm{X}$. In particularly, we set that the dimension of latent representation $k$ equals the number of hash bits in the final binary hash code $\bm{u}_i$. Furthermore, each column of $\bm{V}$, \ie, $\bm{v}_i\in\mathbb{R}^{k}$, can derive $\bm{u}_i$ by
\begin{equation}
\label{eq:hash}
\bm{u}_i = \mathrm{sgn}({\tanh(\bm{v}_i)})\,.
\end{equation}
Meanwhile, regarding $\bm{v}_i$, we also consider to reconstruct its input $\bm{x}_i$ by a decoder as a counterpart of the encoder. Therefore, on one hand, such a reconstruction paradigm can reduce and further distill high-level semantic cues in the attribute space $\mathcal{V}$. While, on the other hand, it can drive the training of \textsc{A$^2$-Net} by preserving the similarity between queried hash codes and database points in terms of hash code learning (cf. Section~\ref{sec:L2H}).

Concretely, the learning objective of unsupervised attribute-guided reconstruction is written as follows:
\begin{equation}\label{eq:WX1}
\min_{\bm{W}} \| \bm{X}- \bm{W}^\top\bm{W}\bm{X}\|_{F}^2 \quad {\rm s.t.}~\bm{W}\bm{X} = \bm{V}'=\tanh(\bm{V})\,,
\end{equation}
where the decoder (\ie, a counterpart of the encoder) is realized by $\bm{W}^\top$ to simplify the network. However, directly minimizing Eq.~\eqref{eq:WX1} with a hard constraint is difficult to optimize. Therefore, we relax the constraint into a soft constraint, and the learning objective can be rewritten as
\begin{equation}\label{eq:WX2}
\min_{\bm{W}} \| \bm{X}- \bm{W}^\top\bm{V}'\|_{F}^2 + \lambda\|\bm{W}\bm{X} - \bm{V}'\|_{F}^2\,.
\end{equation}
%The optimization details of $\bm{W}$ are presented in Section~\ref{sec:optim}.

\subsubsection{Attribute-Specific Feature Decorrelation}\label{sec:ASFD}

By conducting the aforementioned unsupervised attribute-guided learning, we can obtain the internal latent vectors $\bm{V}'$ as the attribute-specific features. In order to both enhance the discriminative ability and remove the redundant correlation among these attribute-specific features, we introduce a feature decorrelation constraint upon $\bm{V}'$, which is formulated by
\begin{equation}\label{eq:featdec}
\min_{\bm{V}'} \| \bm{V}'\bm{V}'^\top- n \bm{I} \|_{F}^2\,,
\end{equation}
where $\bm{I}$ is the identity matrix and $n$ is the batch size. Such a feature decorrelation constraint is preferable to construct independent features and reduce redundant information. Therefore, based on both unsupervised attribute-guided reconstruction and attribute-specific feature decorrelation, the final learned hash codes are expected to be both attribute-aware and hash-bit independent.

\subsubsection{Hash Code Learning}\label{sec:L2H}

%In the hash code learning module, we follow~\cite{qingyuanAAAI18,powernips15} to perform an asymmetric hash learning based on the obtained attribute-specific features $\bm{v}_i$. In concretely, two hash functions, \ie, $h(\cdot)$ and $g(\cdot)$, are employed to learn two different binary codes for the same training sample, which is formulated as

In the following, we conduct the hash code learning based on the obtained attribute-specific features. Assume that we have $n$ query data points which are denoted as $\{\bm{q}_i\}_{i=1}^n$, as well as $m$ database points which are denoted as $\{\bm{v}_j\}_{j=1}^m$. Note that, both $\bm{q}_i$ and $\bm{v}_i$ are belonging to the attribute space $\mathcal{V}$. By following Eq.~\eqref{eq:hash}, the corresponding binary hash codes can be obtained via
\begin{eqnarray}
\label{eq:hash1} \bm{u}_i = \mathrm{sgn}({\tanh(\bm{q}_i)})\,,\\
\label{eq:hash2} \bm{z}_j = \mathrm{sgn}({\tanh(\bm{v}_j)})\,,
\end{eqnarray}
where $\bm{u}_i, \bm{z}_j \in \{-1,+1\}^k$. The goal of our hash code learning is to learn binary hash codes for both query points and database points from $\{\bm{q}_i\}_{i=1}^n$, $\{\bm{v}_j\}_{j=1}^m$, and the pairwise supervised information, \ie, $\bm{S}\in \{-1,+1\}^{n\times m}$. To preserve the pairwise similarity, we adopt the $\ell_2$ loss between the supervised information (\emph{aka} similarity) and the inner product of query-database point binary code pairs. It can be formulated as follows:
\begin{eqnarray}
\label{eq:L2H}
& \min_{\bm{U}, \bm{Z}} \sum_{i=1}^n \sum_{j=1}^m \left( \bm{u}_i^\top \bm{z}_j - k S_{ij}\right)^2 \nonumber\\
& {\rm s.t.}\quad \bm{U}\in \{-1,+1\}^{n\times k}, \bm{Z}\in \{-1,+1\}^{m\times k}\,,
\end{eqnarray}
where $S_{ij}$ is the pairwise similarity label in $\bm{S}$.

Overall, we get the final objective of the \textsc{A$^2$-Net} model by considering Eq.~\eqref{eq:WX2}, Eq.~\eqref{eq:featdec} and Eq.~\eqref{eq:L2H} together as follows:
\begin{align}
\min_{\bm{W},\Theta} \mathcal{L}(\mathcal{I}) &= \| \bm{X}- \bm{W}^\top\bm{V}'\|_{F}^2 + \lambda\|\bm{W}\bm{X} - \bm{V}'\|_{F}^2 \nonumber\\
&+ \alpha \| \bm{V}'\bm{V}'^\top- n \bm{I} \|_{F}^2 + \beta \sum_{i=1}^n \sum_{j=1}^m \left( \bm{u}_i^\top \bm{z}_j - k S_{ij}\right)^2 \,,
\end{align}
where $\lambda$, $\alpha$ and $\beta$ are hyper-parameters as the trade-off.

In practice, during training, it might be only available a set of database points $\{\bm{v}_j\}_{j=1}^m$ without query points. Thus, we randomly sample $n$ data points from database to construct a query set, and denote the indices of all the database points as $\Gamma$ with the indices of the query set as $\Omega$. Additionally, because we cannot back-propagate the gradient to $\Theta$ due to the $\rm sgn(\cdot)$ function, we omit the $\rm sgn(\cdot)$ function and only apply $\tanh(\cdot)$ for relaxation in Eq.~\eqref{eq:hash2} of the whole optimization process. Therefore, the optimization formulation of \textsc{A$^2$-Net} can be rewritten with only database points $\{\bm{v}_j\}_{j=1}^m$ for training as:
\begin{align}\label{eq:final}
\min_{\bm{W},\Theta} \mathcal{L}(\mathcal{I}) &= \| \bm{X}- \bm{W}^\top\bm{V}'\|_{F}^2 + \lambda\|\bm{W}\bm{X} - \bm{V}'\|_{F}^2 \nonumber\\
&+ \alpha \| \bm{V}'\bm{V}'^\top- n \bm{I} \|_{F}^2 \nonumber \\
&+ \beta \sum_{i\in\Omega} \sum_{j\in\Gamma} \left( {\tanh(\bm{W}\cdot F(\mathcal{I}_i;\Theta))}^\top \bm{z}_j - k S_{ij}\right)^2 \,.
\end{align}
For optimization, our \textsc{A$^2$-Net} does not require complicated two-stage learning algorithms, \eg, the alternative optimization strategy. In experiments, we employ the back-propagation algorithm and follow~\cite{qingyuanAAAI18} to train the whole \textsc{A$^2$-Net} model in a unified end-to-end manner.

\subsection{Enhancing Model's Self-Consistency}\label{sec:enhanceSC}

When obtaining attribute-aware internal latent representations of the input images, we employ an encoder-decoder structure as a reconstruction paradigm. Beyond acquiring the implicit representations, such a structure is actually an explicit manifestation of the principle of \emph{self-consistency}~\cite{Yima2022}.

The self-consistency principle~\cite{Yima2022} encourages that an autonomous intelligent system seeks a most self-consistent model for observations of the external world by minimizing the internal discrepancy between the observed and the regenerated. That is to say, the intelligent system should be able to regenerate the distribution of the observed data from the compressed representation to the point that itself cannot distinguish internally despite its best effort. In the contrary, as a consequence, we conjecture that, in the absence of self-consistency, the model will exhibit a tendency to simplicity bias~\cite{simpBiasNIPs20,OODcvpr22} according to the information bottleneck framework~\cite{openblackEntropy2021,deepIBtheoryWS,IB1999} in deep neural networks.

\subsubsection{Preliminary Empirical Studies of Simplicity Bias on Deep Hashing}

Simplicity bias (SB)~\cite{simpBiasNIPs20} is recently observed from neural networks trained with stochastic gradient descent, which shows the networks rely preferentially on few simple predictive features while ignoring more complex predictive features. The SB phenomenon partially explains the lack of robustness and generalization of deep neural networks in various vision problems, \eg, out of distribution~\cite{OODcvpr22}, confidence calibration~\cite{calibICML}, etc. In order to validate and explicitly observe the simplicity bias on deep hashing, we hereby firstly conduct a series of preliminary experiments.

\begin{figure}[t!]
	\centering
	{\includegraphics[width=0.7\columnwidth]{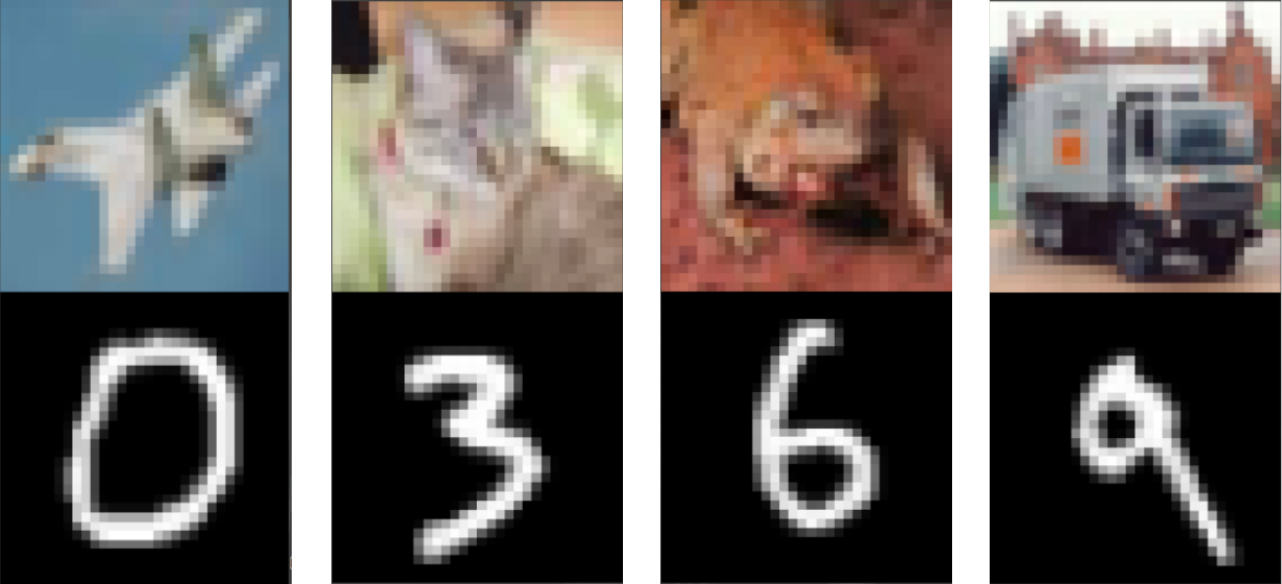}}
	\caption{\small Examples of the \emph{MNIST-CIFAR} dataset~\cite{simpBiasNIPs20}.}
	\label{fig:pre_exp_oriimg}
\end{figure}

Concretely, we follow the \emph{MNIST-CIFAR} dataset introduced in~\cite{simpBiasNIPs20} to perform deep hashing. More specifically, each image of the \emph{MNIST-CIFAR} dataset is constructed by vertically concatenating \emph{MNIST}~\cite{mnist} images onto \emph{CIFAR-10}~\cite{cifar} images. The dataset contains 10 classes, where each class is a combination of a unique \emph{MNIST} class and \emph{CIFAR-10} class, cf. Figure~\ref{fig:pre_exp_oriimg}. It is apparent to see that the \emph{MNIST} part contains much simpler patterns than the \emph{CIFAR} part.

\begin{figure}[t!]
	\centering
	\subfloat[\small {Activations by \textsc{Baseline}}]  {\includegraphics[width=0.7\columnwidth]{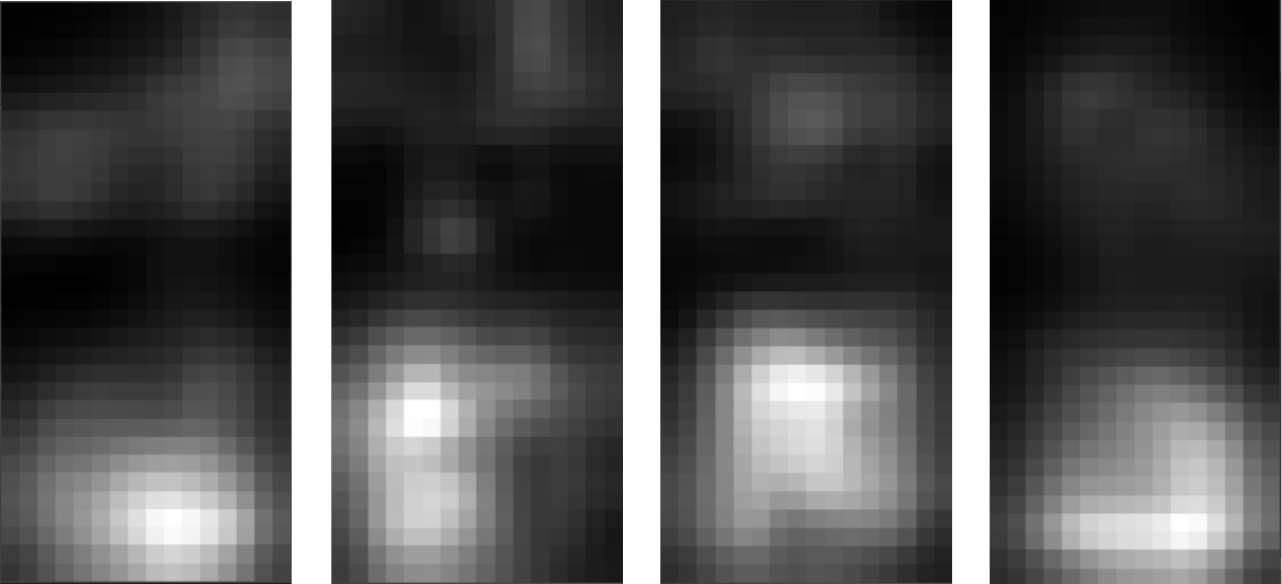} \label{fig:act1}}\\
	\subfloat[\small {Activations by our \textsc{A$^2$-Net}}]  {\includegraphics[width=0.7\columnwidth]{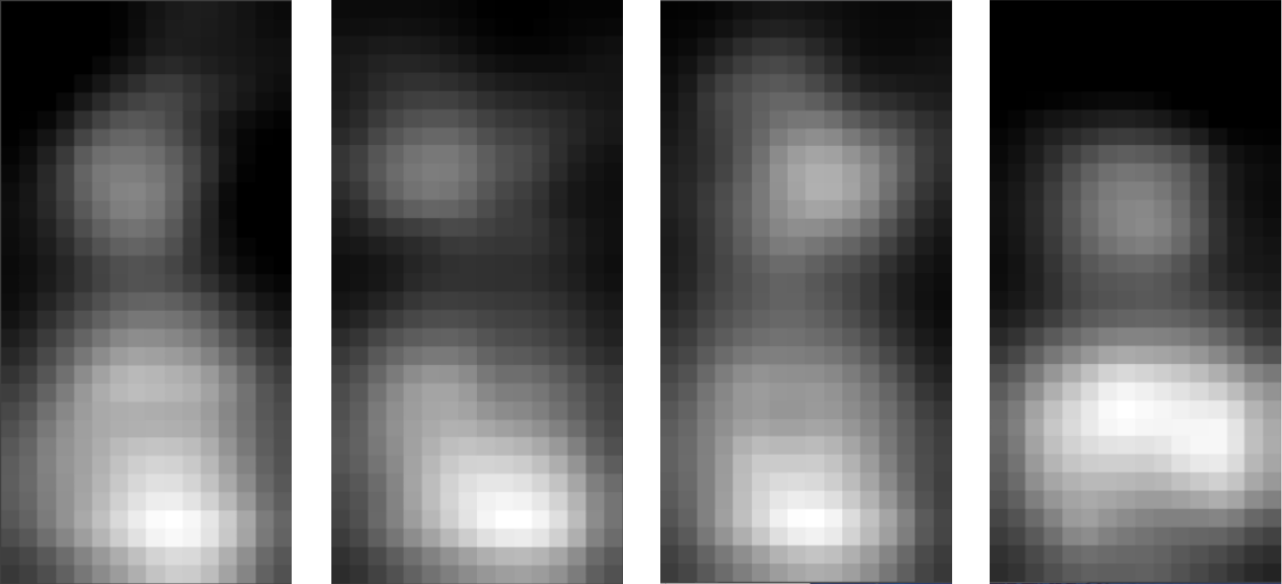} \label{fig:act2}}\\
	\subfloat[\small {Activations by our \textsc{A$^2$-Net$^{++}$}}]  {\includegraphics[width=0.7\columnwidth]{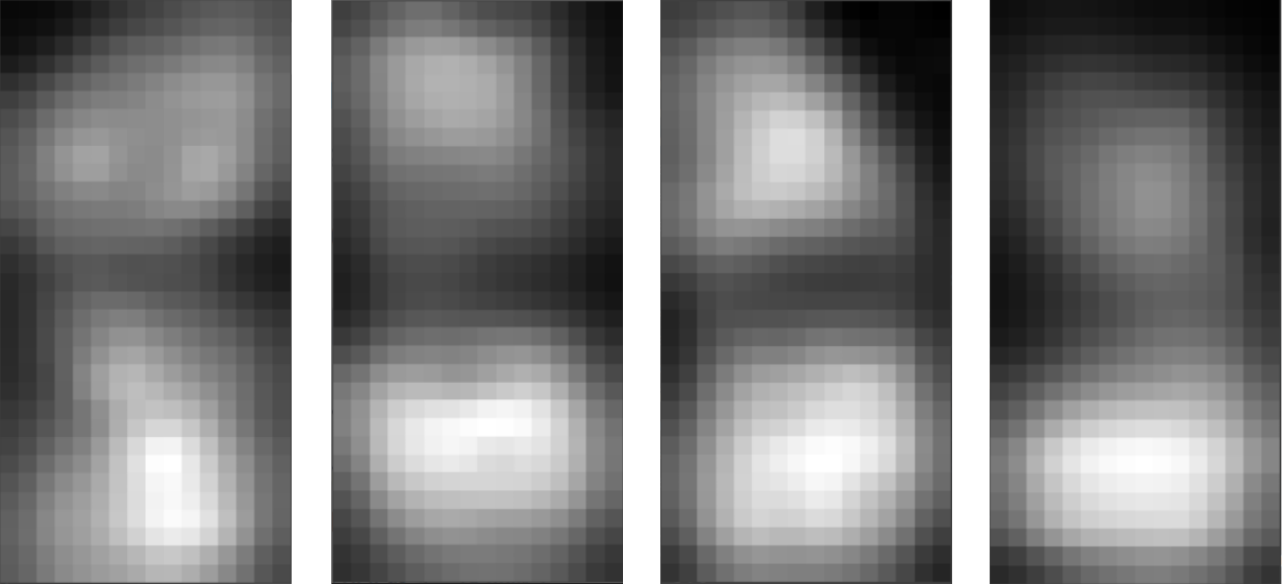} \label{fig:act3}}
	\caption{\small Activations obtained by \textsc{Baseline} with vanilla hashing losses and our proposed models with self-consistency designs.}
	\label{fig:pre_exp_act}
\end{figure}

In our preliminary experiments, we employ the ResNet-32~\cite{resnet16} model as the backbone. Thus, the baseline model is ResNet-32 by equipping with a vanilla learning-to-hash loss to keep the pairwise similarity (cf. Eqn.~\eqref{eq:L2H}), which is denoted as ``\textsc{Baseline}''. We also follow the training/test splits in~\cite{simpBiasNIPs20} to train both \textsc{Baseline} and our \textsc{A$^2$-Net}. Until models convergence, we visualize the activations of input \emph{MNIST-CIFAR} images in the test set (also Figure~\ref{fig:pre_exp_oriimg}). As the activations shown in Figure~\ref{fig:act1}, it can be clearly observe that, for \textsc{Baseline}, the \emph{MNIST} parts are activated while the \emph{CIFAR} parts are partially or nearly completely ignored. This observation indicates SB indeed happens in deep hashing because in this dataset digit patterns of \emph{MNIST} are simpler than the object patterns of \emph{CIFAR}.

To further inspect the SB in deep hashing quantitatively, we leave the training data of \emph{MNIST-CIFAR} intact, while proposing different setups of test data, \ie, \emph{MNIST-CIFAR}, \emph{MNIST-only}, and \emph{CIFAR-only}. \emph{MNIST-only}/\emph{CIFAR-only} means that we only use the \emph{MNIST}/\emph{CIFAR} part for test. The results are reported in the column of \textsc{Baseline} in Table~\ref{table:preMNCI}. As presented, compared with the testing accuracy on \emph{MNIST-CIFAR}, retrieval accuracy on \emph{MNIST-only} has a slight drop, as the network focused mainly on the \emph{MNIST} part. However, the test data of \emph{CIFAR-only} shows a large overall accuracy reduction, which confirms the existence of SB in deep hash learning again.

%=======================table1=============================
\begin{table}[t]%[htbp]
\centering
\small
\setlength{\tabcolsep}{4.5pt}
\caption{Preliminary results of retrieval accuracy (\% mAP) on the \emph{MNIST-CIFAR} dataset.}
\vspace{-1em}
\begin{tabular}{c|c|c|c|c}
\toprule
Test data                           & \# bits & \textsc{Baseline} & \textsc{A$^2$-Net} & \textsc{A$^2$-Net$^{++}$} \\
\hline
\multirow{4}{*}{\emph{MNIST-CIFAR}} & 12      & 99.69             & 99.70              & 99.70            \bigstrut[t]         \\
                                    & 24      & 99.68             & 99.71              & 99.73                     \\
                                    & 32      & 99.71             & 99.71              & 99.72                     \\
                                    & 48      & 99.73             & 99.74              & 99.75                     \\
\hline
\multirow{4}{*}{\emph{MNIST-only}}  & 12      & 99.55             & 99.58              & 99.52                     \\
                                    & 24      & 99.60             & 99.64              & 99.59                     \\
                                    & 32      & 99.36             & 99.37              & 99.32                     \\
                                    & 48      & 99.52             & 99.52              & 99.51                     \\
\hline
\multirow{4}{*}{\emph{CIFAR-only}}  & 12      & 21.05             & 21.23              & 24.50                     \\
                                    & 24      & 21.33             & 22.06              & 25.67                     \\
                                    & 32      & 21.66             & 22.67              & 26.40                     \\
                                    & 48      & 21.33             & 23.54              & 29.40                    \\	
\bottomrule
\end{tabular}%
\label{table:preMNCI}%
\end{table}%

On the other side, regarding the results of our \textsc{A$^2$-Net}, we can find that our model alleviates the SB issue from both qualitative and quantitative aspects. Specifically, as shown in Figure~\ref{fig:act2}, the activations of \textsc{A$^2$-Net} on the \emph{CIFAR} part are significantly enhanced. Meanwhile, the retrieval accuracy of \textsc{A$^2$-Net} on \emph{CIFAR-only} also obtains consistent improvements for all the different lengths of hash codes.

\begin{figure}[t!]
\centering
%\vspace{0.12em}
	{\includegraphics[width=0.99\columnwidth]{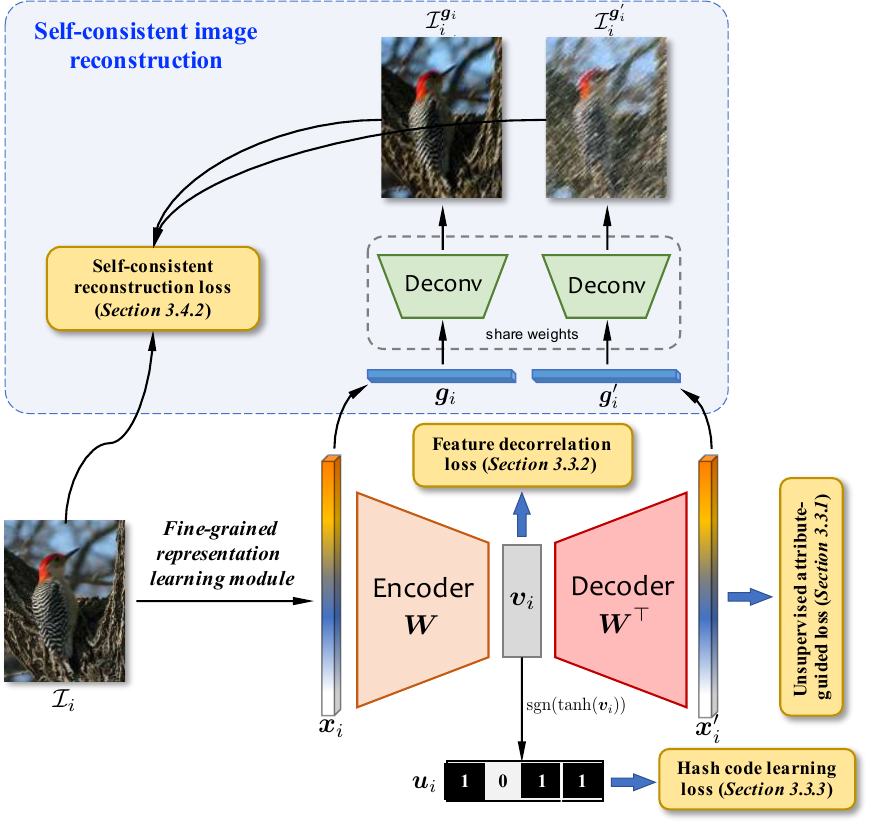}}
\caption{Illustration of the advanced \textsc{A$^2$-Net$^{++}$} model, which has an additional self-consistent image reconstruction modules over \textsc{A$^2$-Net}, and can still be trained in an end-to-end manner.}
\label{fig:a2net++}
\end{figure}

\subsubsection{Advanced \textsc{A$^2$-Net$^{++}$} Model}\label{sec:imageRec}

Combined with the preliminary empirical results and the self-consistency principle argument in~\cite{Yima2022,openblackEntropy2021}, we basically confirm our conjecture about the advantage of our encoder-decoder reconstruction structure in \textsc{A$^2$-Net}. To further alleviate the simplicity bias and then improve the retrieval accuracy, we aim to enhance the model's self-consistency and propose the advanced \textsc{A$^2$-Net$^{++}$} model.

In fact, a more natural way to realize self-consistency is to reconstruct the original image $\mathcal{I}_i$ itself, rather than reconstructing the holistic feature representation $\bm{x}_i$. It could be expected to encourage more comprehensive features to combat with the simplicity bias in deep hashing. Motivated by this, we add two additional reconstructions over $\mathcal{I}_i$ on the basic \textsc{A$^2$-Net}, which is detailedly illustrated in Figure~\ref{fig:a2net++}.

More specifically, based on $\bm{x}_i\in\mathbb{R}^{d}$, we first obtain a more compact feature $\bm{g}_i\in\mathbb{R}^{d'}$ by performing $\bm{W}'\bm{x}_i$, where $\bm{W}'\in\mathbb{R}^{d'\times d} $ is a linear transformation and $d'\ll d$. Then, a deconvolutional network~\cite{deconvCVPR} $\Psi_{\rm Deconv}(\cdot;\Pi)$ is applied upon $\bm{g}_i$, which is expected to reconstruct the input raw image $\mathcal{I}_i$ with its parameter $\Pi$. We denote the reconstructed output w.r.t. $\bm{g}_i$ as $\mathcal{I}^{\bm{g}_i}_i$. In the similar way, we can also get the reconstructed $\mathcal{I}^{\bm{g}'_i}_i$ derived from $\bm{g}'_i$ w.r.t. $\bm{x}'_i$. Regarding $\mathcal{I}^{\bm{g}_i}_i$ and $\mathcal{I}^{\bm{g}'_i}_i$, we minimize the internal discrepancy between their input $\mathcal{I}_i$ and themselves by performing a self-consistent reconstruction loss, which is as follows.
\begin{align}
\min_{\bm{W}', \Pi} &\| \mathcal{I}^{\bm{g}_i}_i- \mathcal{I}_i\|_{F}^2 + \|\mathcal{I}^{\bm{g}'_i}_i- \mathcal{I}_i\|_{F}^2 \nonumber \\
=\min_{\bm{W}', \Pi} &\| \Psi_{\rm Deconv}(\bm{g}_i;\Pi)- \mathcal{I}_i\|_{F}^2 + \|\Psi_{\rm Deconv}(\bm{g}'_i;\Pi)- \mathcal{I}_i\|_{F}^2\nonumber \\
=\min_{\bm{W}', \Pi}& \| \Psi_{\rm Deconv}(\bm{W}'\bm{x}_i;\Pi)- \mathcal{I}_i\|_{F}^2 \nonumber \\ +& \|\Psi_{\rm Deconv}(\bm{W}'\bm{x}'_i;\Pi)- \mathcal{I}_i\|_{F}^2 \,.
\end{align}
Such a dual reconstruction structure of $\bm{g}_i$ and $\bm{g}'_i$ is actually to better enhance the self-consistency nature of the model, while also better constraining the intermediate features, \eg, $\bm{x}_i$ and $\bm{x}'_i$, learned in the previous modules.

Therefore, by incorporating Eq.~\eqref{eq:final}, the final learning objective of our advanced \textsc{A$^2$-Net$^{++}$} model is formulated as:
\begin{align}\label{eq:final2}
\min_{\bm{W},\Theta,\bm{W}',\Pi} \mathcal{L}(\mathcal{I}) &= \| \bm{X}- \bm{W}^\top\bm{V}'\|_{F}^2 + \lambda\|\bm{W}\bm{X} - \bm{V}'\|_{F}^2 \nonumber\\
&+ \alpha \| \bm{V}'\bm{V}'^\top- n \bm{I} \|_{F}^2 \nonumber \\
&+ \beta \sum_{i\in\Omega} \sum_{j\in\Gamma} \left( {\tanh(\bm{W}\cdot F(\mathcal{I}_i;\Theta))}^\top \bm{z}_j - k S_{ij}\right)^2 \nonumber \\
&+\eta\| \Psi_{\rm Deconv}(\bm{W}'\bm{x}_i;\Pi)- \mathcal{I}_i\|_{F}^2 \nonumber\\
&+\eta\|\Psi_{\rm Deconv}(\bm{W}'\bm{x}'_i;\Pi)- \mathcal{I}_i\|_{F}^2 \,,
\end{align}
where $\eta$ is the trade-off parameter of the enhanced self-consistent reconstruction losses. It notes that, enhancing models's self-consistency for \textsc{A$^2$-Net$^{++}$} does not affect the ease of end-to-end training of \textsc{A$^2$-Net}.

Additionally, we apply \textsc{A$^2$-Net$^{++}$} on the \emph{MNIST-CIFAR} dataset to continue the preliminary experiments, and show the qualitative visualization and quantitative results in Figure~\ref{fig:act3} and Table~\ref{table:preMNCI}, respectively. It is clearly to find that \textsc{A$^2$-Net$^{++}$} can further combat simplicity bias based on its enhanced self-consistency design (cf. Figure~\ref{fig:a2net++}).

\subsection{Out-of-Sample Extension}

After training, our learned model can be applied for generating binary codes for query points including unseen query points in the training phase. Specifically, we can use the following equation to generate the binary code for $\mathcal{I}_q$:
\begin{equation}
\bm{u}_q = \mathrm{sgn}({\tanh(\bm{W}\cdot F(\mathcal{I}_q;\Theta) )}) \,.
\end{equation}

\section{Experiments}\label{sec:experiments}

In this section, we present the implementation details, empirical settings, main results, ablation studies and also some interesting qualitative analyses of the learned attribute-aware binary hash codes.

\subsection{Datasets}

Beyond evaluating the retrieval accuracy on fine-grained image retrieval tasks on fine-grained benchmarks, we further involve generic image retrieval datasets for comparisons to justify the effectiveness of our models.

\subsubsection{Fine-Grained Image Retrieval Datasets}

We conduct experiments on five fine-grained benchmark datasets, \ie, \textit{CUB200-2011}~\cite{WahCUB200_2011}, \textit{Aircraft}~\cite{airplanes}, \textit{Food101}~\cite{food101}, \textit{NABirds}~\cite{nabirds15} and \textit{VegFru}~\cite{vegfru}. Specifically, \textit{CUB200-2011} is one of the most popular fine-grained datasets. It contains 11,788 bird images from 200 bird species and is officially split into 5,994 images for training and 5,794 images for test. \textit{Aircraft} contains 10,000 images spanning 100 aircraft models with 3,334 for training, 3,333 for validation and 3,333 for test. For large-scale datasets, \textit{Food101} contains 101 kinds of food with 101,000 images, where for each class, 250 test images are checked manually for correctness while 750 training images still contain a certain amount of noises. \textit{NABirds} has 48,562 images of North American birds with 555 sub-categories, where 23,929 for training and 24,633 for test. \textit{VegFru} is another large-scale fine-grained dataset covering 200 kinds of vegetables and 92 kinds of fruit with 29,200 for training, 14,600 for validation and 116,931 for test.

Furthermore, the \emph{Clothing} dataset~\cite{dualawarericcv15} is also used for evaluating fine-grained retrieval accuracy in the cross-domain scenario, as well as quantitatively performing comparisons in terms of attribute-level matching for our attribute-aware models. This dataset consists of cross-scenario image pairs of online shopping images and corresponding offline user photos with fine-grained clothing attributes. The number of online images is about 450,000 with additional 90,000 offline counterparts collected. Each image has about 5-9 semantic attribute categories, with more than a hundred possible attribute values. The training and test splits in our experiments are followed~\cite{dualawarericcv15}.

%=======================table2=============================
\begin{table*}[t]%[htbp]
\centering
\small
\caption{Comparisons of retrieval accuracy (\% mAP) on five fine-grained benchmark datasets. For fair comparisons, the backbone of these methods in this table is ResNet-50. Best results are marked in bold.}
\vspace{-1em}
\begin{tabular}{c|c|rrrrrr|cc}
\toprule
Datasets & \multicolumn{1}{c|}{\# bits} & \multicolumn{1}{c}{ITQ} & \multicolumn{1}{c}{SDH} & \multicolumn{1}{c}{DPSH} & \multicolumn{1}{c}{HashNet} & \multicolumn{1}{c}{ADSH} & \multicolumn{1}{c|}{ExchNet} & \multicolumn{1}{c}{\textbf{\textsc{A$^2$-Net}}} & \multicolumn{1}{c}{\textbf{\textsc{A$^2$-Net$^{++}$}}} \\
\hline
\multirow{4}[2]{*}{\textit{CUB200-2011}} & 12 & 6.80 & 10.52 & 8.68  & 12.03 & 20.03 & 25.14 & {33.83}& \textbf{37.83} \bigstrut[t]\\
& 24 & 9.42 & 16.95 & 12.51 & 17.77 & 50.33 & 58.98 & {61.01} & \textbf{71.73}\\
& 32 & 11.19 & 20.43 & 12.74 & 19.93 & 61.68 & 67.74 & {71.61}& \textbf{78.39} \\
& 48 & 12.45 & 22.23 & 15.58 & 22.13 & 65.43 & 71.05 & {77.33} & \textbf{82.71}\\
\hline
\multirow{4}[2]{*}{\textit{Aircraft}} & 12 & 4.38 & 4.89 & 8.74 & 14.91 & 15.54 & 33.27 & {40.00}& \textbf{57.53} \bigstrut[t]\\
& 24 & 5.28 & 6.36 & 10.87 & 17.75 & 23.09 & 45.83 & {63.66}& \textbf{73.45} \\
& 32 & 5.82 & 6.90 & 13.54 & 19.42 & 30.37 & 51.83 & {72.51}& \textbf{81.59} \\
& 48 & 6.05 & 7.65 & 13.94 & 20.32 & 50.65 & 59.05 & {81.37}& \textbf{86.65} \\
\hline
\multirow{4}[2]{*}{\textit{Food101}} & 12 & 6.46 & 10.21 & 11.82 & 24.42 & 35.64 & 45.63 & {46.44}& \textbf{54.51} \bigstrut[t]\\
& 24 & 8.20 & 11.44 & 13.05 & 34.48 & 40.93 & 55.48 & {66.87}& \textbf{81.46} \\
& 32 & 9.70 & 13.36 & 16.41 & 35.90 & 42.89 & 56.39 & {74.27}& \textbf{82.92} \\
& 48 & 10.07 & 15.55 & 20.06 & 39.65 & 48.81 & 64.19 & {82.13} & \textbf{83.66}\\
\hline
\multirow{4}[2]{*}{\textit{NABirds}} & 12 & 2.53 & 3.10 & 2.17 & 2.34 & 2.53 & 5.22 & ~~{8.20} & ~~\textbf{8.80}\bigstrut[t]\\
& 24 & 4.22 & 6.72 & 4.08 & 3.29 & 8.23 & 15.69 & {19.15} & \textbf{22.65}\\
& 32 & 5.38 & 8.86 & 3.61 & 4.52 & 14.71 & 21.94 & {24.41}& \textbf{29.79} \\
& 48 & 6.10 & 10.38 & 3.20 & 4.97 & 25.34 & 34.81 & {35.64} & \textbf{42.94}\\
\hline
\multirow{4}[2]{*}{\textit{VegFru}} & 12 & 3.05 & 5.92 & 6.33 & 3.70 & 8.24 & 23.55 & {25.52}& \textbf{30.54} \bigstrut[t]\\
& 24 & 5.51 & 11.55 & 9.05 & 6.24 & 24.90 & 35.93 & {44.73}& \textbf{60.56} \\
& 32 & 7.48 & 14.55 & 10.28 & 7.83 & 36.53 & 48.27 & {52.75} & \textbf{73.38}\\
& 48 & 8.74 & 16.45 & 9.11 & 10.29 & 55.15 & 69.30 & {69.77}& \textbf{82.80} \\	
\bottomrule
\end{tabular}%
\label{table:results_fg}%
\end{table*}%

\subsubsection{Generic Image Retrieval Datasets}

\emph{NUS-WIDE}~\cite{NUSWIDE} and \emph{COCO}~\cite{cocoeccvMS} are two popular generic image benchmark datasets, which are employed as test beds for generic image retrieval evaluation. Concretely, \emph{NUS-WIDE}~\cite{NUSWIDE} consists of 269,648 web images associated with tags. It is a multi-label dataset where each image might be annotated with multi-labels. We choose images from the 21 most frequent categories for evaluation by following~\cite{qingyuanAAAI18,csqCVPR20}. \emph{COCO}~\cite{cocoeccvMS} is another multi-label dataset, containing 82,783 training, 40,504 validation images which belong to 91 categories. For the training image set, we discard the images which have no category information. Following the protocols of these datasets in generic deep hashing~\cite{qingyuanAAAI18,csqCVPR20,onelossforall}, two images will be defined as a ground-truth neighbor (similar pair) if they share at least one common label.

\subsection{Comparison Methods and Implementation Details}\label{sec:expdetails}

We hereby introduce the comparison methods and implementation details of our models.

\subsubsection{Comparison Methods}
	
In experiments of fine-grained hashing, we compare our proposed models to the following competitive hashing methods, \ie, ITQ~\cite{gong2012iterative}, SDH~\cite{shen2015supervised}, DPSH~\cite{li2015feature}, HashNet~\cite{cao2017hashnet}, and ADSH~\cite{qingyuanAAAI18}. Among them, DPSH, HashNet and ADSH are deep learning based methods. Furthermore, we also compare the results of our \textsc{A$^2$-Net} and \textsc{A$^2$-Net$^{++}$} with state-of-the-arts of fine-grained hashing methods, including ExchNet~\cite{exchnet}. Additionally, other fine-grained hashing methods, \ie, DSaH~\cite{jin2020deep} and FISH~\cite{tipxinshunTIP}, also achieved good retrieval accuracy. However, as their empirical settings are different from other mainstream fine-grained hashing methods, for fair comparisons, we strictly control empirical settings the same as those of~\cite{jin2020deep} and \cite{tipxinshunTIP} and compare the results of our models with their results in the appendices.

For the comparisons of generic hashing, we further involve two recent state-of-the-art methods, \ie, CSQ~\cite{csqCVPR20} and OrthoCos+BN~\cite{onelossforall}, in experiments on the \emph{NUS-WIDE} and \emph{COCO} generic image retrieval datasets.

\subsubsection{Implementation Details}\label{sec:implem}
	
For fair comparisons of fine-grained hashing, we follow the efficient training setting in ExchNet~\cite{exchnet}. Concretely, for \textit{CUB200-2011}, \textit{Aircraft} and \textit{Food101}, we sample 2,000 images per epoch, while 4,000 samples are randomly selected for \textit{NABirds}, \textit{VegFru} and \textit{Clothing}. For the training details, regarding the backbone model, we can choose any network structure as the base network for the fine-grained representation learning module. Following~\cite{exchnet}, ResNet-50~\cite{resnet16} is employed in experiments. For generic hashing, 4,000 samples are randomly selected for \emph{NUS-WIDE} and \emph{COCO}. The backbone model employs the CNN-F model~\cite{bmvcdetailsZA} by following~\cite{qingyuanAAAI18}.

For \textit{CUB200-2011}, \textit{Aircraft}, \textit{Food101}, \textit{NABirds} and \textit{VegFru}, we follow~\cite{exchnet} to perform such settings upon our proposed models. The total number of training epochs is $20$, and the number of batch size is set as $16$. Different from~\cite{exchnet}, our model only requires a smaller iteration number to converge. Specifically, for these datasets containing less than 20,000 training images, the iteration number $T_{\rm max}$ is $60$, and the learning rate is divided by $10$ at the $50^{\rm th}$ iteration. For other datasets, $T_{\rm max}$ is set as $70$, and the learning rate is divided by $10$ at the $60^{\rm th}$ iteration. 

For \emph{NUS-WIDE}, \emph{COCO} and \emph{Clothing}, we adopt the training settings from~\cite{qingyuanAAAI18,wujunhashingijcai16}. Concretely, regarding \emph{NUS-WIDE} and \emph{COCO}, the total number of training epochs is $3$, and $T_{\rm max}$ is set as $50$. Regarding \emph{Clothing}, the total number of training epochs is $10$, and $T_{\rm max}$ is set as $15$. The number of batch size for these three datasets is set as $16$. The learning rate is $10^{-4}$.

For all datasets, the image resolution is $224\times 224$. The optimizer of our models is standard mini-batch stochastic gradient descent with the weight decay as $10^{-4}$. The number of attention guidance equals the number of hash bits. The value of $d'$ in \textsc{A$^2$-Net$^{++}$} is set to $1024$. The hyper-parameters, \ie, $\lambda$, $\alpha$ and $\beta$ in Eq.~\eqref{eq:final2}, are set as $1$, $\frac{1}{n\times k}$ and $\frac{12}{k}$, respectively. The value of $\eta$ is $0.5$ for the \emph{MNIST-CIFAR} dataset, while being $0.1$ for other datasets.%All experiments are conducted with a GeForce RTX 2080 Ti GPU.

\begin{figure*}[p]
\centering
	\subfloat[\emph{CUB200-2011}]  {\includegraphics[width=0.71\textwidth]{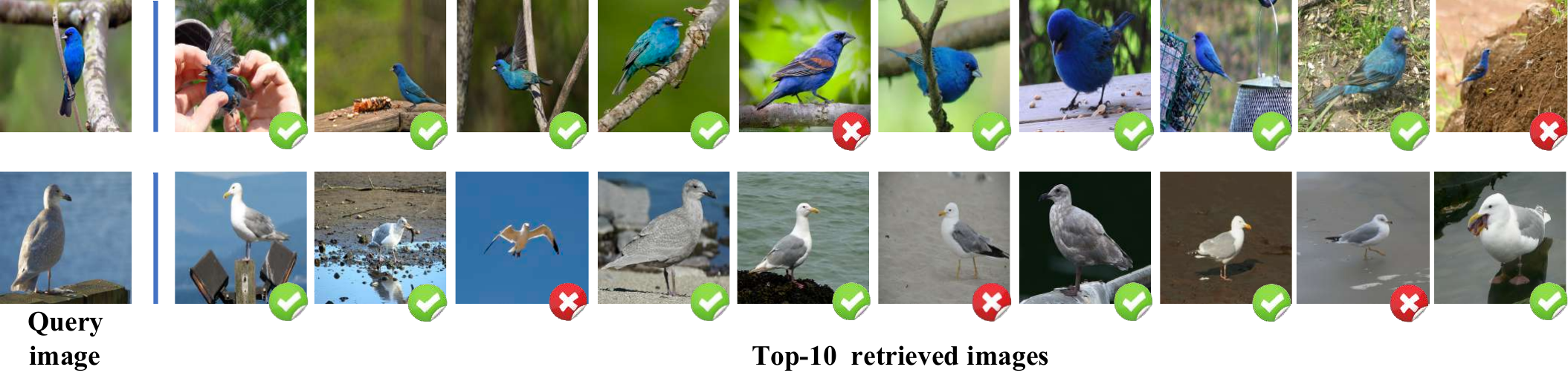} }\\
	\subfloat[\emph{Aircraft}]  {\includegraphics[width=0.71\textwidth]{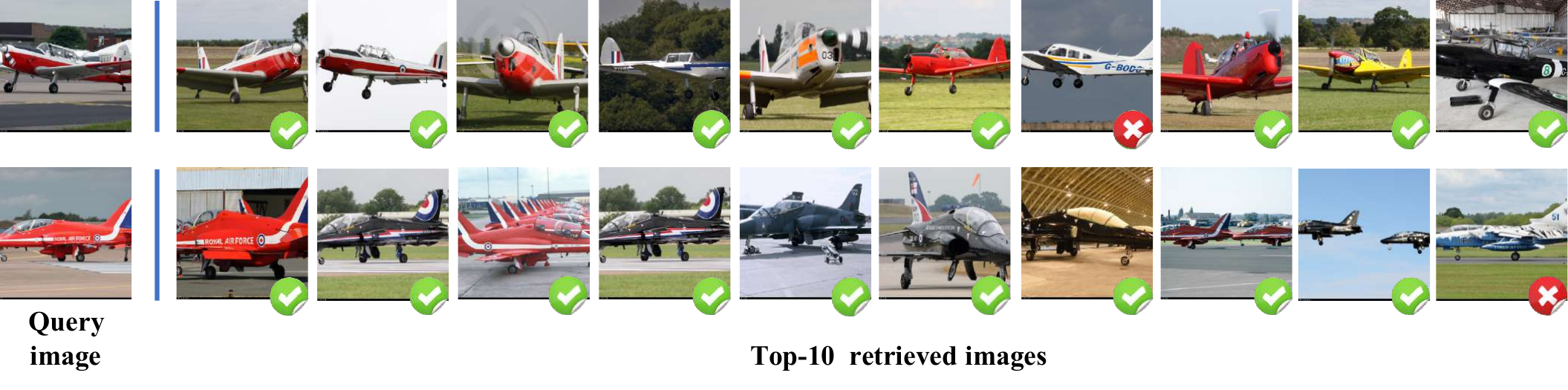} }\\
	\subfloat[\emph{Food101}]  {\includegraphics[width=0.71\textwidth]{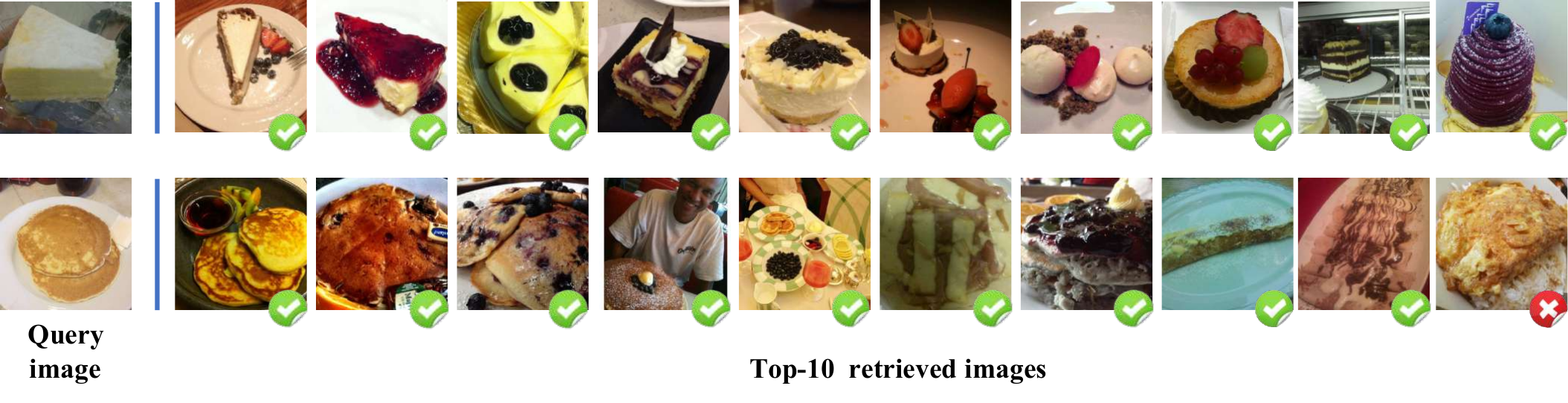} }\\
	\subfloat[\emph{NABirds}]  {\includegraphics[width=0.71\textwidth]{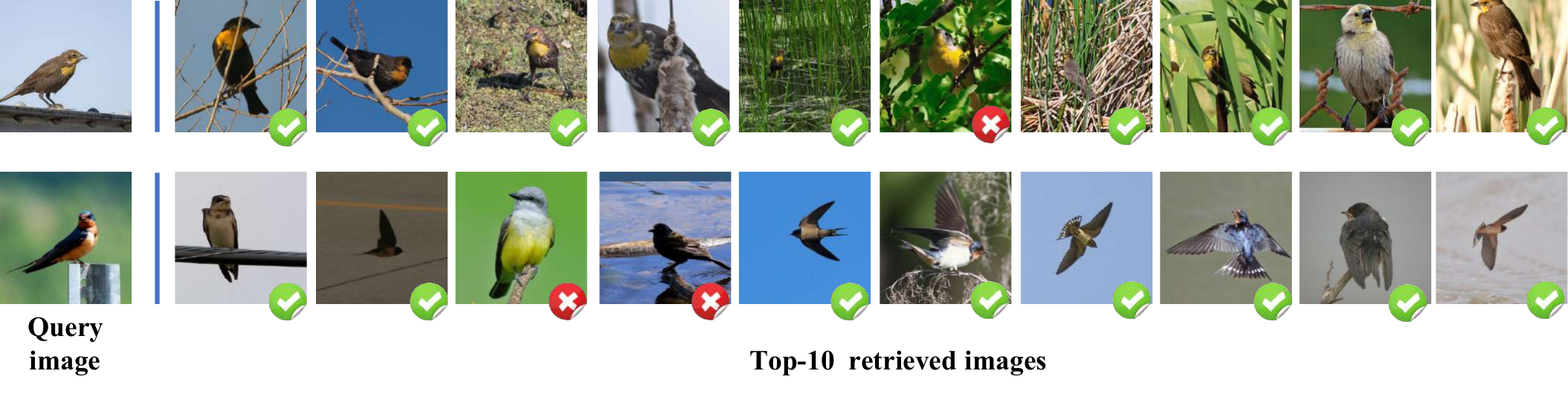} }\\
	\subfloat[\emph{VegFru}]  {\includegraphics[width=0.71\textwidth]{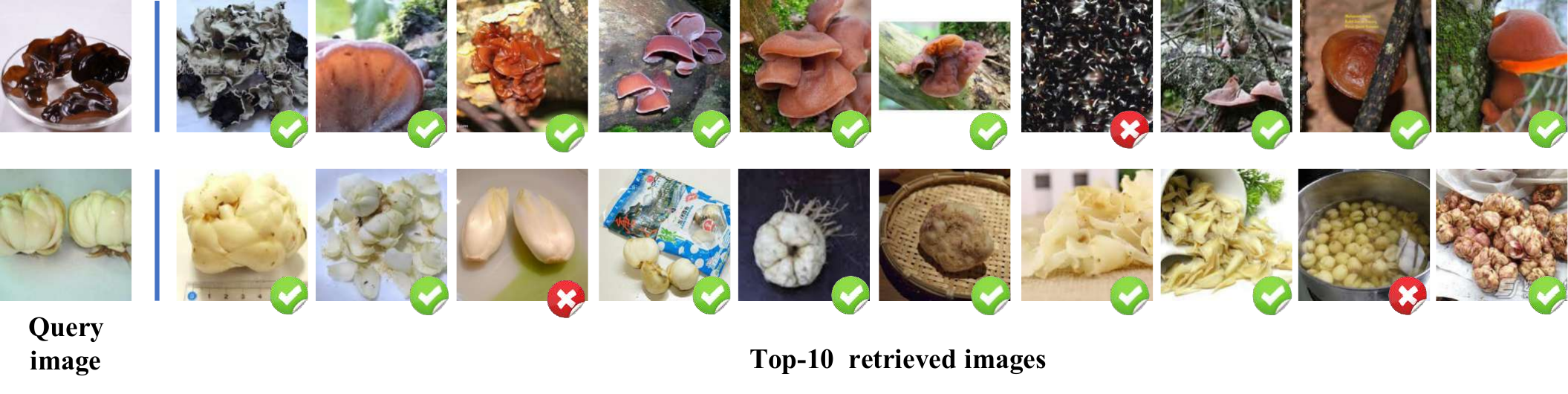} }\\
	\subfloat[\emph{Clothing}]  {\includegraphics[width=0.71\textwidth]{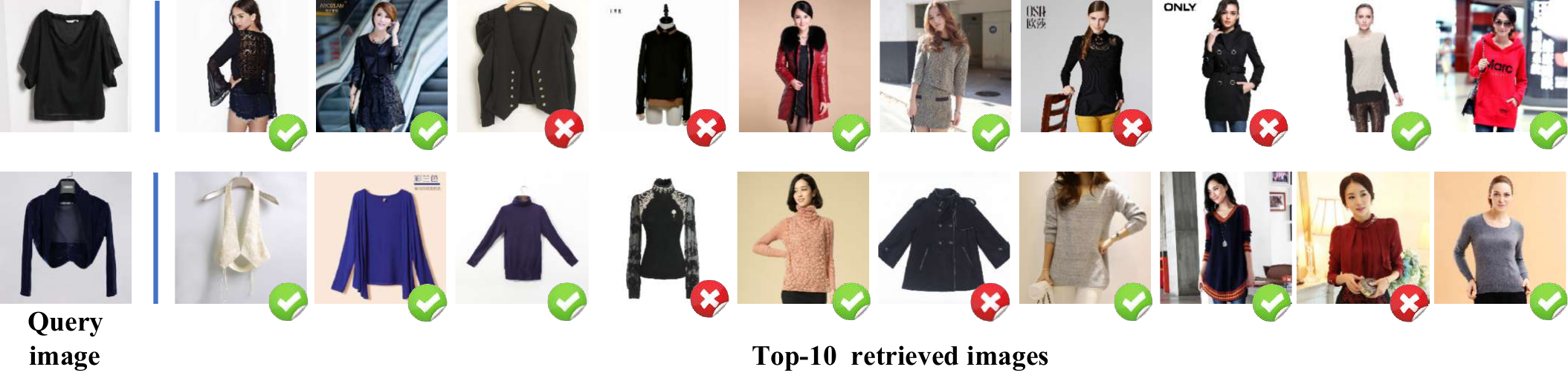} \label{fig:clothingretri}}
%\vspace{-0.5em}
\caption{Examples of top-10 retrieved images on six fine-grained datasets of 48-bit hash codes by our \textsc{A$^2$-Net$^{++}$}.}
\label{fig:retri_res}
\end{figure*}

\subsection{Main Results}

We present the main results from the following five aspects: 1) On fine-grained image datasets for validating the effectiveness; 2) On generic image datasets for investigating the universality; 3) In zero-shot fine-grained retrieval setting for proving the generalization ability; 4) In cross-domain fine-grained retrieval setting for justifying the robustness; and 5) Time and space complexity during inference.

% \textsc{A$^2$-Net$^{++}$}

%=======================table3=============================
\begin{table*}[t]%[htbp]
\centering
\small
\caption{Comparisons of retrieval accuracy (\% mAP) on two generic image datasets. For fair comparisons, the backbone of these methods in this table is CNN-F. Best results are marked in bold.}
\vspace{-1em}
\begin{tabular}{c|c|ccccccc|cc}
\toprule
Datasets & \multicolumn{1}{c|}{\# bits} & \multicolumn{1}{c}{ITQ} & \multicolumn{1}{c}{SDH} & \multicolumn{1}{c}{DPSH} & \multicolumn{1}{c}{HashNet} & \multicolumn{1}{c}{ADSH} & \multicolumn{1}{c}{CSQ} & \multicolumn{1}{c|}{OrthoCos+BN} & \multicolumn{1}{c}{\textbf{\textsc{A$^2$-Net}}} & \multicolumn{1}{c}{\textbf{\textsc{A$^2$-Net$^{++}$}}} \\
\hline
\multirow{4}[2]{*}{\textit{NUS-WIDE}} & 12 & 71.43 & 76.46 & 79.41 & 76.20 & 84.00 & 77.64 & 76.85 & 82.95 & \textbf{84.93} \bigstrut[t]\\
& 24 & 73.61 & 79.98 & 82.49 & 78.54 & 87.84 & 78.92 & 79.64 & 87.57 & \textbf{88.90}\\
& 32 & 74.57 & 80.17 & 83.51 & 79.31 & 89.51 & 80.29 & 81.37 & 88.13 & \textbf{89.87} \\
& 48 & 75.53 & 81.24 & 84.42 & 79.76 & 90.55 & 80.52 & 82.01 & 89.43 & \textbf{90.68}\\
\hline
\multirow{4}[2]{*}{\textit{COCO}} & 12 & 63.38 & 69.54 & 74.61 & 76.49 & 83.88 & 79.40 & 76.00 & 86.42 & \textbf{87.84} \bigstrut[t]\\
& 24 & 63.26 & 70.78 & 76.67 & 77.85 & 85.90 & 80.84 & 78.63 & 88.14 & \textbf{88.76} \\
& 32 & 63.08 & 71.15 & 77.29 & 78.38 & 86.33 & 82.16 & 80.35 & 88.97 & \textbf{89.91} \\
& 48 & 63.38 & 71.64 & 77.77 & 79.17 & 86.51 & 82.26 & 81.60 & 89.28 & \textbf{90.27} \\
\bottomrule
\end{tabular}%
\label{table:results_generic}%
\end{table*}%

%=======================table4=============================
\begin{table*}[t]%[htbp]
\centering
\small
\caption{Results in terms of retrieval accuracy (\% mAP) with the zero-shot retrieval protocol~\cite{ZShash} on three fine-grained datasets. For fair comparisons, the backbone of these methods in this table is ResNet-50. Best results are marked in bold.}
\vspace{-1em}
\begin{tabular}{c|c|ccccc|cc}
\toprule
Datasets & \multicolumn{1}{c|}{\# bits} & \multicolumn{1}{c}{HashNet} & \multicolumn{1}{c}{ADSH} & \multicolumn{1}{c}{CSQ} & \multicolumn{1}{c}{OrthoCos+BN} & \multicolumn{1}{c|}{ExchNet}& \multicolumn{1}{c}{\textbf{\textsc{A$^2$-Net}}} & \multicolumn{1}{c}{\textbf{\textsc{A$^2$-Net$^{++}$}}} \\
\hline
\multirow{4}[2]{*}{\textit{CUB200-2011}} & 12 & 13.02 & 22.41 & 20.41 & 23.30 & 23.53 &23.71 & \textbf{23.89} \bigstrut[t]\\
& 24 & 22.58 & 39.02 & 31.51 & 38.82 & 39.51 &39.67 & \textbf{41.32}\\
& 32 & 32.45 & 46.98 & 41.55 & 46.30 & 48.13 &51.33 & \textbf{52.75} \\
& 48 & 42.02 & 52.91 & 48.32 & 51.89 & 53.77 &58.47 & \textbf{62.71}\\
\hline
\multirow{4}[2]{*}{\textit{Aircraft}} & 12 & 16.54 & 21.84 & 22.34 & 22.14 & 23.03 &23.45 & \textbf{24.14} \bigstrut[t]\\
& 24 & 32.72 & 45.16 & 43.85 & 44.93 & 47.90 &51.62 & \textbf{53.87} \\
& 32 & 39.44 & 50.95 & 49.49 & 49.38 & 53.22 &58.91 & \textbf{60.38} \\
& 48 & 39.75 & 55.31 & 51.66 & 53.49 & 56.01 &59.63 & \textbf{61.59} \\
\hline
\multirow{4}[2]{*}{\textit{Food101}} & 12 & 17.89 & 22.89 & 23.54 & 23.18 & 24.69 &26.32 & \textbf{27.98} \bigstrut[t]\\
& 24 & 36.54 & 42.65 & 41.74 & 42.01 & 43.11 &45.25 & \textbf{47.85} \\
& 32 & 41.06 & 46.98 & 46.63 & 46.10 & 48.25 &52.74 & \textbf{53.27} \\
& 48 & 41.85 & 50.31 & 47.72 & 50.69 & 51.87 &55.97 & \textbf{56.30} \\
\bottomrule
\end{tabular}%
\label{table:results_zs}%
\end{table*}%

%=======================table5=============================
\begin{table*}[t]%[htbp]
\centering
\small
\caption{Results in terms of retrieval accuracy (\% mAP) within cross-domain scenarios on the \emph{Clothing} dataset~\cite{dualawarericcv15}. For fair comparisons, the backbone of these methods in this table is ResNet-50. Best results are marked in bold.}
\vspace{-1em}
\begin{tabular}{c|c|cccccc|cc}
\toprule
Datasets & \multicolumn{1}{c|}{\# bits} & \multicolumn{1}{c}{ITQ} & \multicolumn{1}{c}{SDH} & \multicolumn{1}{c}{DPSH} & \multicolumn{1}{c}{HashNet} & \multicolumn{1}{c}{ADSH} & \multicolumn{1}{c|}{ExchNet} & \multicolumn{1}{c}{\textbf{\textsc{A$^2$-Net}}} & \multicolumn{1}{c}{\textbf{\textsc{A$^2$-Net$^{++}$}}} \\
\hline
\multirow{4}[2]{*}{\textit{Clothing}} & 12 & 32.18 & 32.67 & 32.03 & 34.06 & 34.82 & 36.38 & 37.71 & \textbf{38.00} \bigstrut[t]\\
& 24 & 32.25 & 33.39 & 32.87 & 33.84 & 35.64 & 37.06 & 38.11 & \textbf{38.98}\\
& 32 & 33.08 & 33.54 & 33.22 & 34.44 & 36.08 & 37.73 & 38.43 & \textbf{39.14} \\
& 48 & 33.91 & 34.32 & 34.15 & 34.51 & 36.33 & 38.24 & 39.04 & \textbf{39.33}\\
\bottomrule
\end{tabular}%
\label{table:results_crossdomain}%
\end{table*}%

\subsubsection{Comparisons on Fine-Grained Image Datasets}

Table~\ref{table:results_fg} presents the mean average precision (mAP) results of fine-grained retrieval on these five aforementioned fine-grained benchmark datasets. For each dataset, we report the results of four hash bits with varying lengths, \ie, 12, 24, 32, and 48, for evaluation. As shown in the table, our proposed \textsc{A$^2$-Net} and \textsc{A$^2$-Net$^{++}$} models consistently outperform the other baseline methods on these datasets significantly. In particular, compared with the state-of-the-art method ExchNet~\cite{exchnet}, our \textsc{A$^2$-Net} (\textsc{A$^2$-Net$^{++}$}) achieves 17.83\% (27.62\%) and 17.88\% (26.53\%) improvements over ExchNet of 24-bit and 32-bit experiments on \emph{Aircraft} and \emph{Food-101}, respectively. Moreover, \textsc{A$^2$-Net} also obtains superior results with an absolute value of about 80\% mAP on \emph{CUB200-2011}, \emph{Aircraft} and \emph{Food101} with 48-bit hash codes. While \textsc{A$^2$-Net$^{++}$} achieves much better results over \textsc{A$^2$-Net} thanks to its simple but effective self-consistency enhancement, cf. Section~\ref{sec:imageRec}. These observations validate the effectiveness of our models, as well as their promising practicality in real-world applications of fine-grained retrieval. Additionally, in Figure~\ref{fig:retri_res}, we illustrate several retrieval results of \textsc{A$^2$-Net$^{++}$} on these fine-grained datasets, which shows that it can retrieve well among multiple subordinate categories when the same species of visual objects with diverse variations appear in different kinds of background. Also, there exist several failure cases, due to extremely subtle visual differences (\eg, caused by different views) between the query image and the returned images.
%which clearly shows our advantages in terms of both retrieval accuracy and consistency.

\subsubsection{Comparisons on Generic Image Datasets}

To further investigate the effectiveness and universality of our models, we conduct experiments on two generic image retrieval datasets, \ie, \emph{NUS-WIDE} and \emph{COCO}. Except these classical and competitive baseline methods, we also compare with two recent state-of-the-arts in the field of generic hashing, \ie, CSQ~\cite{csqCVPR20} and OrthoCos+BN~\cite{onelossforall}. As reported in Table~\ref{table:results_generic}, our {\textsc{A$^2$-Net}} achieves comparable or slightly better results over strong baselines, including ADSH, CSQ and OrthoCos+BN. While, our {\textsc{A$^2$-Net$^{++}$}} consistently outperforms all methods by a large margin. The results prove that our models can also handle generic image hashing tasks well, rather than being limited to fine-grained hashing tasks.

\subsubsection{Results of Zero-Shot Fine-Grained Retrieval}

In the aforementioned experiments, we follow the traditional supervised hashing protocol, \ie, supervised retrieval protocol where queries and database have identical classes~\cite{cao2017hashnet,ZShash}. In order to further investigate the generalization ability of our models, the challenging zero-shot retrieval protocol~\cite{ZShash} is also deployed, where queries and database have \emph{different classes}. We compare several representative methods of both fine-grained hashing and generic hashing, and report the results on three fine-grained datasets in Table~\ref{table:results_zs}. As shown, our proposed models are still the optimal solution under the zero-shot retrieval setting, especially for {\textsc{A$^2$-Net$^{++}$}}. Its retrieval accuracy is consistently and significantly better than other competing methods.

%=======================table=============================
\begin{table}[t]%[htbp]
\centering
\small
\caption{Time and space complexity comparisons with baseline methods. Our \textsc{A$^2$-Net} and \textsc{A$^2$-Net$^{++}$} models exhibit similar computational complexity during inference. ``$\uparrow$'' (``$\downarrow$'') denotes that the higher (lower) the better.}
\vspace{-1em}
\begin{tabular}{c|ccc}
\toprule
Methods        & WCtime ($\downarrow$)  & Speedup ($\uparrow$)        & Memory ($\downarrow$)  \\
\hline
Linear        & 9481.03 & ~~~~~~~~1$\times$      & 207.2MB \\
BallTree~\cite{balltree}      & ~~236.23  & ~~40.13$\times$  & ~~28.1MB  \\
KDTree~\cite{kdtree}        & ~~~~70.16   & 135.13$\times$ & ~~28.8MB  \\
PQ~\cite{pq}            & ~~~~43.49   & 217.99$\times$ & 524.5KB \\
\hline
\textbf{Ours} & ~~~~\textbf{39.42}   & \textbf{240.50$\times$} & \textbf{380.2KB}\\
\bottomrule
\end{tabular}%
\label{table:timespace}%
\end{table}%

%=======================table6=============================
\begin{table*}[t]%[htbp]
\centering
\scriptsize
\setlength{\tabcolsep}{2.3pt}
\caption{Retrieval accuracy (\% mAP) with incremental components of the proposed {\textsc{A$^2$-Net$^{++}$}} model. The results of \textsc{A$^2$-Net$^{++}$} are reported in the penultimate column. We also present {\textsc{A$^2$-Net}}'s results in the last column for clearly comparisons. Note that, the ``+ Image reconstruction'' column has three variants of equal status. Best results are marked in bold.}
\vspace{-1em}

\begin{tabular}{c|c|ccccccc|c}
\toprule
\multirow{3}{*}{Datasets}          & \multirow{3}{*}{\# bits} & Vanilla   & + Attention           & + Feature             & \multicolumn{1}{c}{+ Image}     & \multicolumn{1}{c}{+ Image} & \multicolumn{1}{c}{+ Image}               & \multicolumn{1}{c|}{+ Feature}             & \multicolumn{1}{c}{\multirow{3}{*}{{\textsc{A$^2$-Net}}}} \\
                                   &                          & Backbone  & (Sec.~\ref{sec:FGRL}) & reconstruction        & \multicolumn{1}{c}{reconstruction}   & \multicolumn{1}{c}{reconstruction}& \multicolumn{1}{c}{reconstruction}         & \multicolumn{1}{c|}{decorrelation}         & \multicolumn{1}{c}{}                                             \\
                                   &                          & (ResNet-50) &                       & (Sec.~\ref{sec:UAGL}) & \multicolumn{1}{c}{(only $\mathcal{I}^{\bm{g}_i}_i$ in Sec.~\ref{sec:imageRec})}& \multicolumn{1}{c}{(only $\mathcal{I}^{\bm{g}'_i}_i$ in Sec.~\ref{sec:imageRec})}& \multicolumn{1}{c}{(both $\mathcal{I}^{\bm{g}_i}_i$/$\mathcal{I}^{\bm{g}'_i}_i$ in Sec.~\ref{sec:imageRec})} & \multicolumn{1}{c|}{(Sec.~\ref{sec:ASFD})} & \multicolumn{1}{c}{}                                             \\
\hline
\multirow{4}[2]{*}{\textit{CUB200-2011}} & 12 & 20.03 & 27.42 & 33.31 &33.42&34.27& 36.13 & \textbf{37.83} & 33.83 \bigstrut[t]\\
& 24 & 50.33 & 58.17 & 60.65 & 60.69&63.15&67.73 & \textbf{71.73} & 61.01 \\
& 32 & 61.68 & 68.24 & 71.28 & 71.36&74.63&77.67 & \textbf{78.39} & 71.61 \\
& 48 & 65.43 & 76.10 & 77.10 & 77.15&79.54&81.83 & \textbf{82.71} & 77.33 \\
\hline
\multirow{4}[2]{*}{\textit{Aircraft}} & 12 & 15.54 & 39.78 & 41.54 & 41.72&54.67&52.06 & \textbf{57.53} & 40.00 \bigstrut[t]\\
& 24 & 23.09 & 61.73 & 64.47 & 64.60&66.24&70.63 & \textbf{73.45} & 63.66 \\
& 32 & 30.37 & 69.34 & 71.09 & 71.18&75.49&79.60 & \textbf{81.59} & 72.51 \\
& 48 & 50.65 & 79.21 & 81.14 & 81.22&82.93&84.52 & \textbf{86.65} & 81.37 \\
\hline
\multirow{4}[2]{*}{\textit{Food101}} & 12 & 35.64 & 41.33 & 45.02 & 45.63&47.50&50.67 & \textbf{54.51} & 46.44 \bigstrut[t]\\
& 24 & 40.93 & 65.07 & 67.49 & 68.04&70.21&74.64 & \textbf{81.46} & 66.87 \\
& 32 & 42.89 & 70.06 & 73.57 & 73.74&74.51&75.73 & \textbf{82.92} & 74.27 \\
& 48 & 48.81 & 78.51 & 81.63 & 81.74&82.33&82.70 & \textbf{83.66} & 82.13 \\
\hline
\multirow{4}[2]{*}{\textit{NABirds}} & 12 & ~~2.34 & ~~5.03 & ~~7.91 & ~~7.93&~~8.13&~~8.46 & ~~\textbf{8.80} & ~~8.20\bigstrut[t]\\
& 24 & ~~8.23 & 16.31 & 18.55 & 18.71&20.09&21.32 & \textbf{22.65} & 19.15 \\
& 32 & 14.71 & 22.34 & 23.87 & 23.96&26.88&28.14 & \textbf{29.79} & 24.41 \\
& 48 & 25.34 & 33.47 & 34.93 & 35.10&36.65&40.04 & \textbf{42.94} & 35.64 \\
\hline
\multirow{4}[2]{*}{\textit{VegFru}} & 12 & ~~8.24 & 22.14 & 24.79 & 24.98&26.43&27.86 & \textbf{30.54} & 25.52 \bigstrut[t]\\
& 24 & 24.90 & 39.69 & 44.21 & 44.76&51.07&55.83 & \textbf{60.56} & 44.73 \\
& 32 & 36.53 & 47.08 & 51.04 & 51.80&63.41&67.98 & \textbf{73.38} & 52.75 \\
& 48 & 55.15 & 66.47 & 68.63 & 69.25&74.68&77.80 & \textbf{82.80} & 69.77 \\	
\bottomrule
\end{tabular}%
\label{table:results_abla}%
\end{table*}%

\subsubsection{Results of Cross-Domain Fine-Grained Retrieval}

Cross-domain fine-grained retrieval is a practical task for many real-world applications, \eg, mobile product image search~\cite{dualawarericcv15}. In experiments, we also consider such a protocol to test the robustness of our proposed models. In this concrete case, we employ the \emph{Clothing} dataset~\cite{dualawarericcv15} for retrieval accuracy evaluation. The corresponding empirical results are presented in Table~\ref{table:results_crossdomain}. Similar observations are obtained with those ones in Table~\ref{table:results_fg}, where our proposed \textsc{A$^2$-Net} and {\textsc{A$^2$-Net$^{++}$}} models significantly outperform other methods in the cross-domain fine-grained retrieval setting. It also reveals that our models could perform well in various practical scenarios with good potential. Also, we show the retrieval results of the \emph{Clothing} dataset in Figure~\ref{fig:clothingretri} for qualitative demonstration.

\subsubsection{Time and Space Complexity During Inference}

Retrieval efficiency plays a vital role in fine-grained hashing. To evaluate the efficiency of our models, we compare them with baseline approaches for approximate nearest neighbor (ANN) search, including two tree-based ANN approaches (\ie, BallTree~\cite{balltree} and KDTree~\cite{kdtree}) and one product quantization based ANN approach (\ie, Product Quantization, PQ~\cite{pq}). These baselines serve as references to assess the performance of our models in terms of wall clock time (WCtime), speedup ratio, and memory cost. In Table~\ref{table:timespace}, we present the results of these comparisons. Our models demonstrate superior performance, achieving the lowest WCtime, highest speedup, and smallest memory cost when compared to the baseline ANN approaches. Notably, our models exhibit a remarkable query speedup of approximately 240 times.

\subsection{Ablation Studies}

In this section, we demonstrate the effectiveness of the crucial components of the proposed {\textsc{A$^2$-Net$^{++}$}} model, \ie, the attention-based fine-grained representation learning component (\emph{aka} ``attention'', cf. Section~\ref{sec:FGRL}), the unsupervised attribute-guided reconstruction component (\emph{aka} ``feature reconstruction'', cf. Section~\ref{sec:UAGL}), the self-consistent image reconstruction component (\emph{aka} ``image reconstruction'', cf. Section~\ref{sec:imageRec}) and the attribute-specific feature decorrelation component (\emph{aka} ``feature decorrelation'', cf. Section~\ref{sec:ASFD}). In the ablation studies, we apply these components incrementally on a vanilla backbone (\ie, ResNet-50). As evaluated in Table~\ref{table:results_abla}, by stacking these components one by one, the retrieval results are steadily improved, which justifies the effectiveness of our proposed components in \textsc{A$^2$-Net$^{++}$}. Besides, the results of \textsc{A$^2$-Net} are presented as well, which can be compared with the results of the ``+ Image reconstruction'' column for clearly observing the large accuracy improvement derived from our novel self-consistency enhancement in Section~\ref{sec:imageRec} over the basic \textsc{A$^2$-Net} model. Furthermore, we also provide more sufficient results of three variants of ``+ Image reconstruction'', \ie, reconstruction with only $\mathcal{I}^{\bm{g}_i}_i$, with only $\mathcal{I}^{\bm{g}'_i}_i$ and with both $\mathcal{I}^{\bm{g}_i}_i$/$\mathcal{I}^{\bm{g}'_i}_i$. It is clear to see that, reconstruction with only $\mathcal{I}^{\bm{g}_i}_i$ brings slightly improvements, and reconstruction with only $\mathcal{I}^{\bm{g}'_i}_i$ improves the retrieval accuracy by a large margin. Reconstruction with both $\mathcal{I}^{\bm{g}_i}_i$/$\mathcal{I}^{\bm{g}'_i}_i$ achieves further improvements over only $\mathcal{I}^{\bm{g}_i}_i$ or $\mathcal{I}^{\bm{g}'_i}_i$, which not only verifies the proposed dual reconstruction is necessary, but also reveals that overcoming simplicity bias by our proposed self-consistent image reconstruction module significantly benefits fine-grained retrieval tasks.

\begin{figure}[t!]
	\centering
	{\includegraphics[width=0.9\columnwidth]{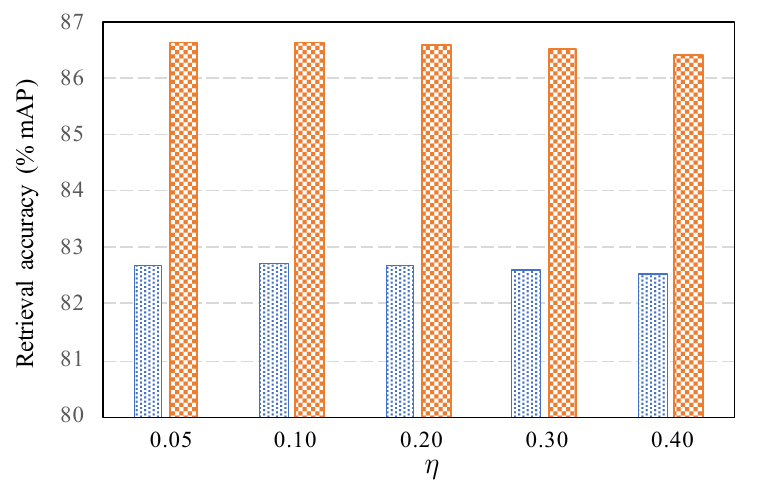}}
	%\vspace{-1em}
	\caption{\small Comparisons of retrieval accuracy (\% mAP) of different values of $\eta$, cf. Eq.~\eqref{eq:final2}. The blue bars represent the results for the \emph{CUB200-2011} dataset, while the orange bars represent the results for the \emph{Aircraft} dataset. All results are based on 48-bit hash codes.}
	\label{fig:eta}
\end{figure}

In addition, the final learning objective of our \textsc{A$^2$-Net$^{++}$} model has four hyper-parameters, including $\lambda$, $\alpha$, $\beta$, and $\eta$, cf. Eq.~\eqref{eq:final2}. Thanks to the robustness of our model, $\lambda$ is fixed to $1$ for all experiments, and $\alpha$ / $\beta$ can be set according to the number of hash bits.\footnote{In the implementation details (cf. Section~\ref{sec:implem}), $\alpha$ and $\beta$ are set to $\frac{1}{n\times k}$ and $\frac{12}{k}$, respectively, where $n$ is the number of training data in a batch and $k$ is the number of hash bits.} Regarding $\eta$, we hereby change its values of $\{0.05, 0.10, 0.20,  0.30, 0.40\}$ for parameter analyses. As depicted in Figure~\ref{fig:eta}, minimal variations can be observed across different values of $\eta$. This finding indicates that our model is not sensitive to parameter changes, highlighting its practical potential.

\begin{figure*}[t!]
	\centering
	\subfloat[\emph{CUB200-2011}]  {\includegraphics[width=0.146\textwidth]{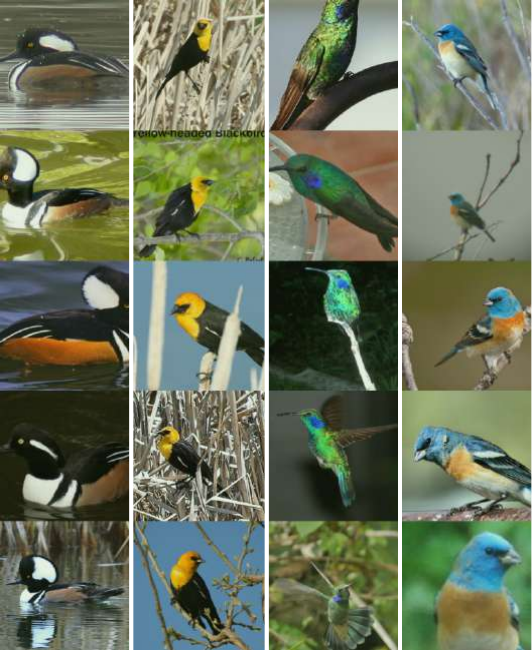} } \quad
	\subfloat[\emph{Aircraft}]  {\includegraphics[width=0.146\textwidth]{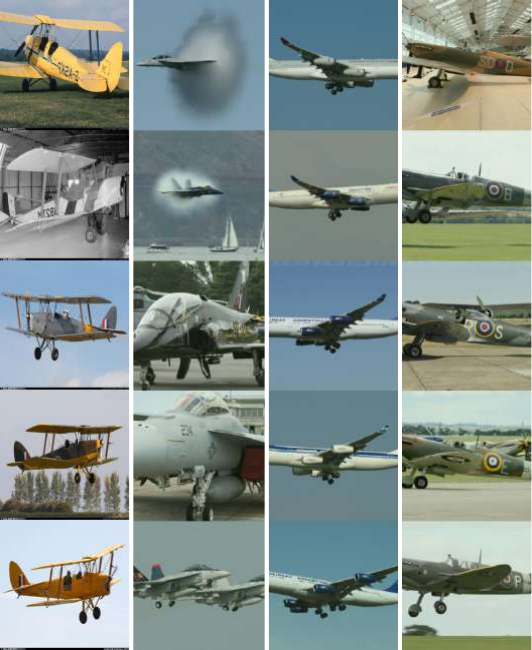} }\quad
	\subfloat[\emph{Food101}]  {\includegraphics[width=0.146\textwidth]{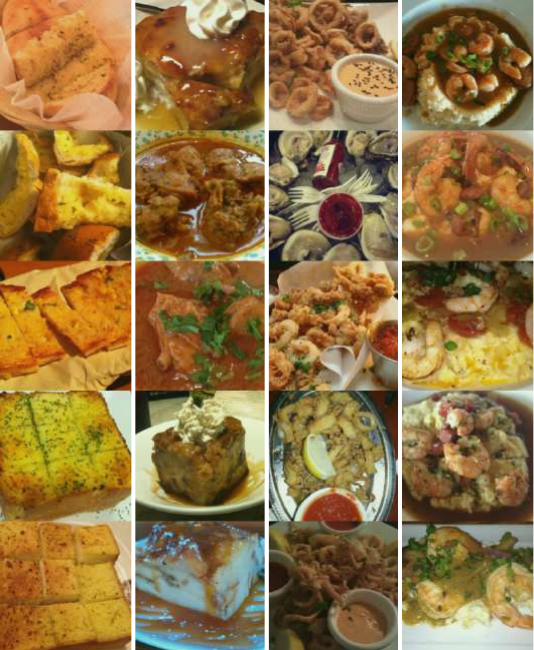} } \quad
	\subfloat[\emph{NABirds}]  {\includegraphics[width=0.146\textwidth]{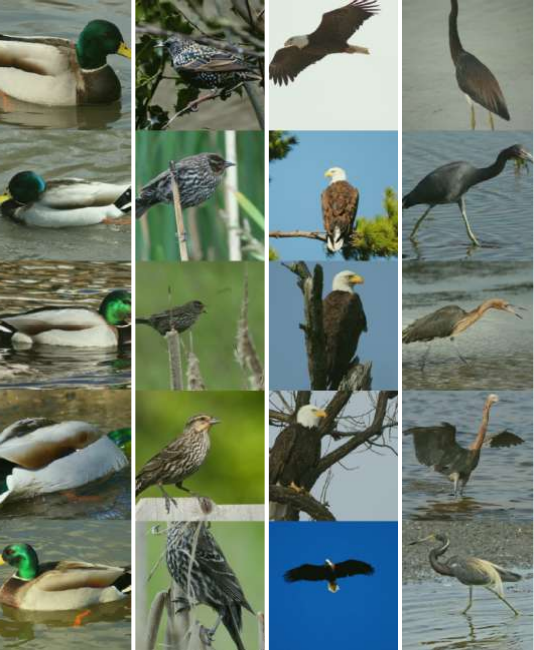} }\quad
	\subfloat[\emph{VegFru}]  {\includegraphics[width=0.146\textwidth]{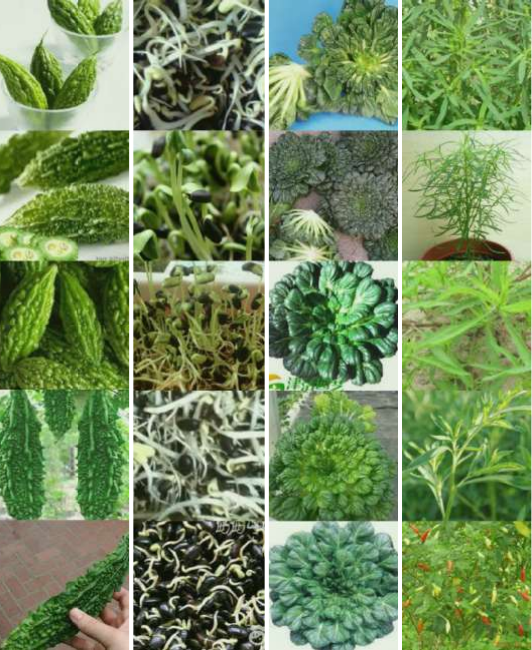} }\quad
	\subfloat[\emph{Clothing}]  {\includegraphics[width=0.146\textwidth]{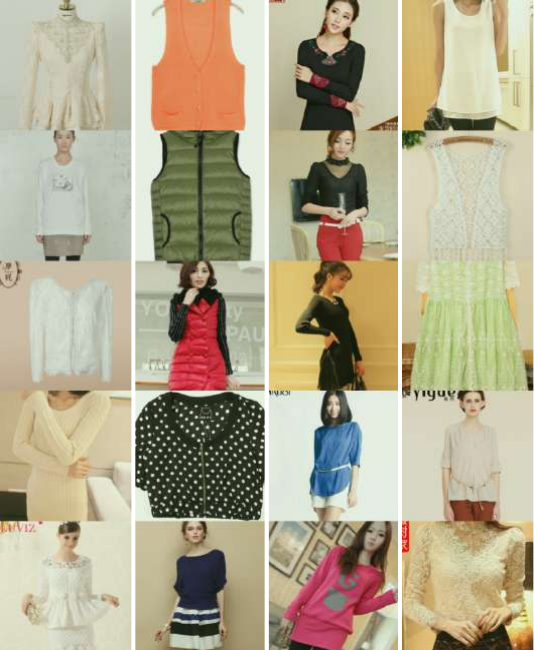} }\quad
%	\vspace{-0.5em}
	\caption{Quality demonstrations of the learned attribute-aware hash codes by the proposed \textsc{A$^2$-Net$^{++}$} model. Each column in each sub-figure can strongly correspond to a certain kind of properties of the fine-grained objects, \eg, ``birds with black and white head in water'', ``double-winged aircrafts'', ``lumpy food'', ``vegetables with uneven surfaces'', ``lace clothing'', etc. (Best viewed in color and zoomed in.)}
	\label{fig:att_visual}
\end{figure*}

\subsection{Attribute-Aware Retrieval Evaluation}

As an important characteristic of our models, we hereby focus on the attribute-aware retrieval evaluation from both qualitative and quantitative perspectives.

\subsubsection{Qualitative Analyses of Attribute-Aware Hash Codes}

We firstly discuss the quality of the learned attribute-aware hash codes $\bm{u}_i$ of \textsc{A$^2$-Net$^{++}$}. After obtaining $\bm{u}_i$, we visualize fine-grained images retrieved by a random single hash bit of $\bm{u}_i$ to demonstrate the strong correspondence between visual attributes and the obtained hash bits. All the six datasets (also including \emph{Clothing}) in experiments are used as examples to illustrate the quality. As observed in Figure~\ref{fig:att_visual}, images of each column have some similar fine-grained object properties, \ie, visual attributes. Indeed, the learned hash codes are apparently attribute-aware, which could provide an explanation of the \textsc{A$^2$-Net$^{++}$}'s success in fine-grained retrieval. Meanwhile, it also offers human-understandable interpretation for such a deep learning based fine-grained hashing method. Moreover, our \textsc{A$^2$-Net} has similar characteristics of attribute awareness~\cite{XSA2Net}.

\subsubsection{Quantitative Results of Attribute Matching}

%=======================table7=============================
\begin{table}[t]%[htbp]
\centering
\small
\caption{Results in terms of NDCG@$20$ for evaluating the attribute-level matching on the \emph{Clothing} dataset. Best results are marked in bold.}
\vspace{-1em}
\begin{tabular}{c|c}
\toprule
Methods                   & NDCG@20                                             \\
\hline
ITQ                       & 0.296   \bigstrut[t]\\
SDH                       & 0.308   \\
DPSH                      & 0.319   \\
HashNet                   & 0.337   \\
ADSH                      & 0.342   \\
ExchNet                   & 0.374   \\
\hline
\textbf{\textsc{A$^2$-Net}}      & 0.396 \bigstrut[t]  \\
\textbf{\textsc{A$^2$-Net$^{++}$}} & \textbf{0.405}  \\
\bottomrule
\end{tabular}%
\label{table:results_attrimat}%
\end{table}%

\begin{figure*}[t!]
	\centering
	\subfloat[\small {\textit{CUB200-2011}}]  {\includegraphics[width=0.193\textwidth]{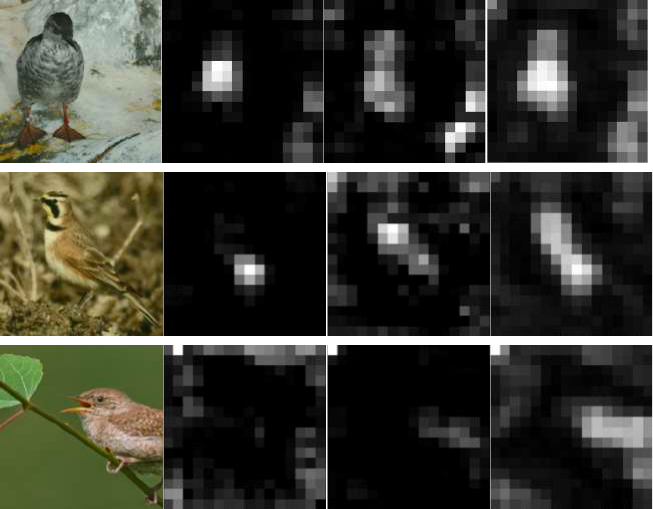} \label{fig:sb1}} \qquad 
	\subfloat[\small {\textit{Aircraft}}]  {\includegraphics[width=0.193\textwidth]{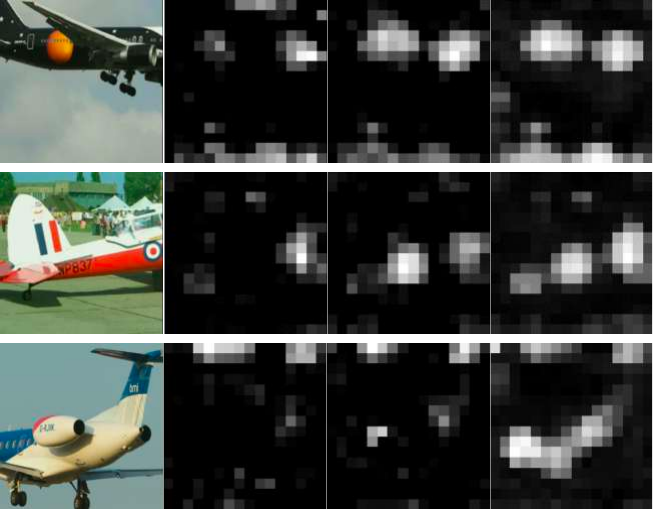} \label{fig:sb2}} \qquad
	\subfloat[\small {\textit{Food101}}]  {\includegraphics[width=0.193\textwidth]{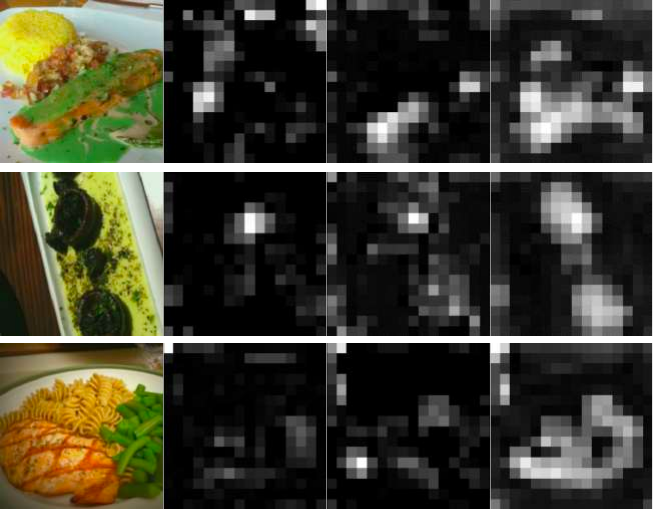} \label{fig:sb3}}  \\
	\subfloat[\small {\textit{NABirds}}]  {\includegraphics[width=0.193\textwidth]{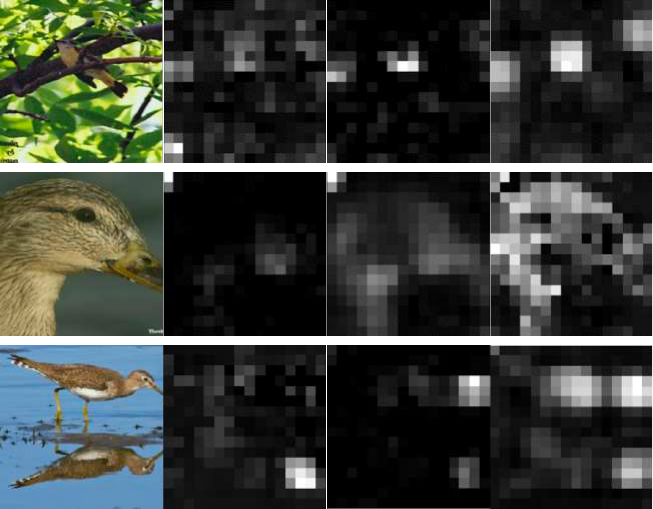} \label{fig:sb4}} \qquad
	\subfloat[\small {\textit{VegFru}}]  {\includegraphics[width=0.193\textwidth]{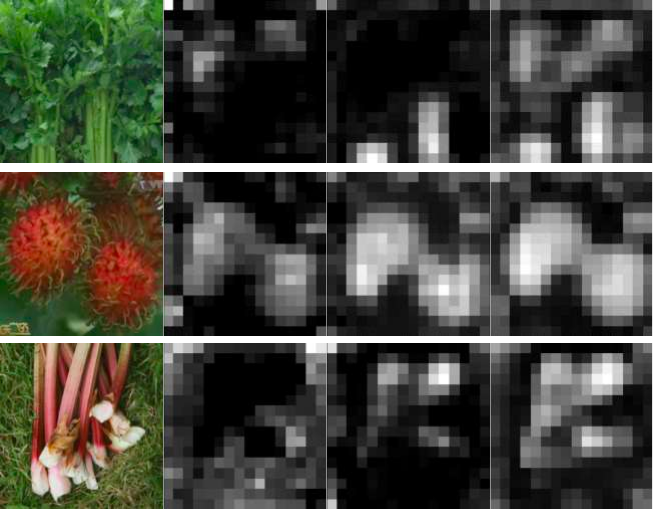} \label{fig:sb5}} \qquad
	\subfloat[\small {\textit{Clothing}}]  {\includegraphics[width=0.193\textwidth]{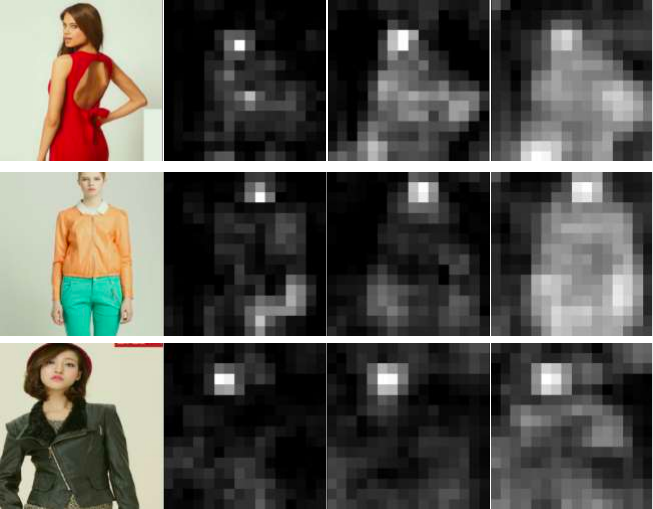} \label{fig:sb6}} 
	\caption{\small Input images and the corresponding activations. In each subfigure, the first column is the input image. The second column presents the activations of ExchNet. The third and forth columns are the activations of our \textsc{A$^2$-Net} and \textsc{A$^2$-Net$^{++}$}.}
	\label{fig:sb_visual}
\end{figure*}

Regarding the quantitative evaluation of our attribute-aware hash codes, we calculate the Normalized Discounted Cumulative Grain (NDCG@$k$) metric~\cite{dualawarericcv15} by performing our models on the \emph{Clothing} dataset. Specifically, NDCG@$k$ equals to $\frac{1}{Z}\sum_{j=1}^{k}\frac{2^{rel(j)}-1}{\log(j+1)}$, where $rel(j)$ is the relevance score of the $j$-th ranked image, and $Z$ is a normalization constant. The relevance score $rel(j)$ of query image and the $j$-th ranked image is the number of their matched attributes divided by the total number of query attributes. The detailed evaluation results in terms of NDCG@$20$ are reported in Table~\ref{table:results_attrimat}, where our models achieve the best score w.r.t strong attribute-level matching. It can quantitatively prove that our learned hash codes contains more similar attributes to the queries, as well as having strong explicit semantic meaning.

\subsection{Visualization of Simplicity Bias on Fine-Grained Retrieval Datasets}

Except for the activations on the \emph{MNIST-CIFAR} dataset shown in Figure~\ref{fig:pre_exp_act}, we further visualize the activations on six fine-grained datasets (also including \emph{Clothing}). As presented in Figure~\ref{fig:sb_visual}, we list the original input images and compare the corresponding activations of ExchNet~\cite{exchnet} and our models. It is apparent to observe that although ExchNet obtains relatively  good quantitative retrieval results of fine-grained hashing (cf. Table~\ref{table:results_fg}), its activations concentrate on some image regions corresponding to simple patterns but overlooking complex ones, which suggests that it still suffers from the pitfall of simplicity bias~\cite{simpBiasNIPs20,huh2021low}. In contrast, thanks to the self-consistency nature of the feature reconstruction process in our \textsc{A$^2$-Net}, it focuses on more (meaningful) image regions corresponding to complex patterns than ExchNet. This enables \textsc{A$^2$-Net} to learn more comprehensive fine-grained patterns and then to learn more discriminative binary hash codes for fine-grained hashing. Regarding \textsc{A$^2$-Net$^{++}$}, further enhancing self-consistency by equipping image reconstruction makes it perform much better than \textsc{A$^2$-Net}. In Figure~\ref{fig:sb_visual}, we can see that basically \emph{all} object information has been completely attended, especially without ignoring the complex patterns, \eg, textures of bird's head, special painting of aircrafts' fuselage, etc.

\section{Conclusion}\label{sec:conc}

In this paper, we proposed attribute-aware hashing networks with self-consistency for dealing with the large-scale fine-grained image retrieval task. Particularly, our models were designed as expected to be efficient, effective and more importantly interpretable. Concretely, by developing an unsupervised attribute-guided reconstruction method based on the obtained appearance-specific visual representation with attention, it can distill attribute-specific vectors in a high-level attribute space. After further performing feature decorrelation upon attribute-specific vectors, their discriminative ability is strengthened for representing a fine-grained object. Then, hash codes can be generated from these attribute-specific vectors and thus became attribute-aware. Through validation experiments of simplicity bias in deep hashing, we considered the model design from the perspective of the self-consistency principle and further enhanced model's self-consistency by equipping an additional image reconstruction path. Both qualitative and quantitative experiments of various datasets and settings justified the superior of our models from different aspects in terms of effectiveness, robustness, generalization ability, interpretation, etc. Furthermore, visual attributes have been shown to be useful in describing visual entities, which motivates us to study identifying both fine-grained and generic categories under challenging scenarios, \eg, novel class discovery, or weakly-supervised categorization/detection, based on our attribute-aware mechanism as the future work.

\section*{Acknowledgment}

The authors would like to thank the editor and the anonymous reviewers for their critical and constructive comments and suggestions. We gratefully acknowledge the support of MindSpore, CANN (Compute Architecture for Neural Networks) and Ascend AI Processor used for this research.

\iffalse
\ifCLASSOPTIONcompsoc
  \section*{Acknowledgments}
\else
  \section*{Acknowledgment}
\fi

\ifCLASSOPTIONcaptionsoff
  \newpage
\fi

\fi

%\section*{Acknowledgments}

%The authors would like to thank the editor and the anonymous reviewers for their constructive comments. 
%This work was supported by National Key R\&D Program of China (2021YFA1001100), National Natural Science Foundation of China under Grant (62272231), Natural Science Foundation of Jiangsu Province of China under Grant (BK20210340), the Fundamental Research Funds for the Central Universities (No. NJ2022028), and CAAI-Huawei MindSpore Open Fund. We gratefully acknowledge the support of MindSpore, CANN (Compute Architecture for Neural Networks) and Ascend AI Processor used for this research.

\bibliographystyle{IEEEtran}
\bibliography{FGIA_survey_full_A2Net}

\begin{IEEEbiography}[{\includegraphics[width=1in,height=1.25in,clip,keepaspectratio]{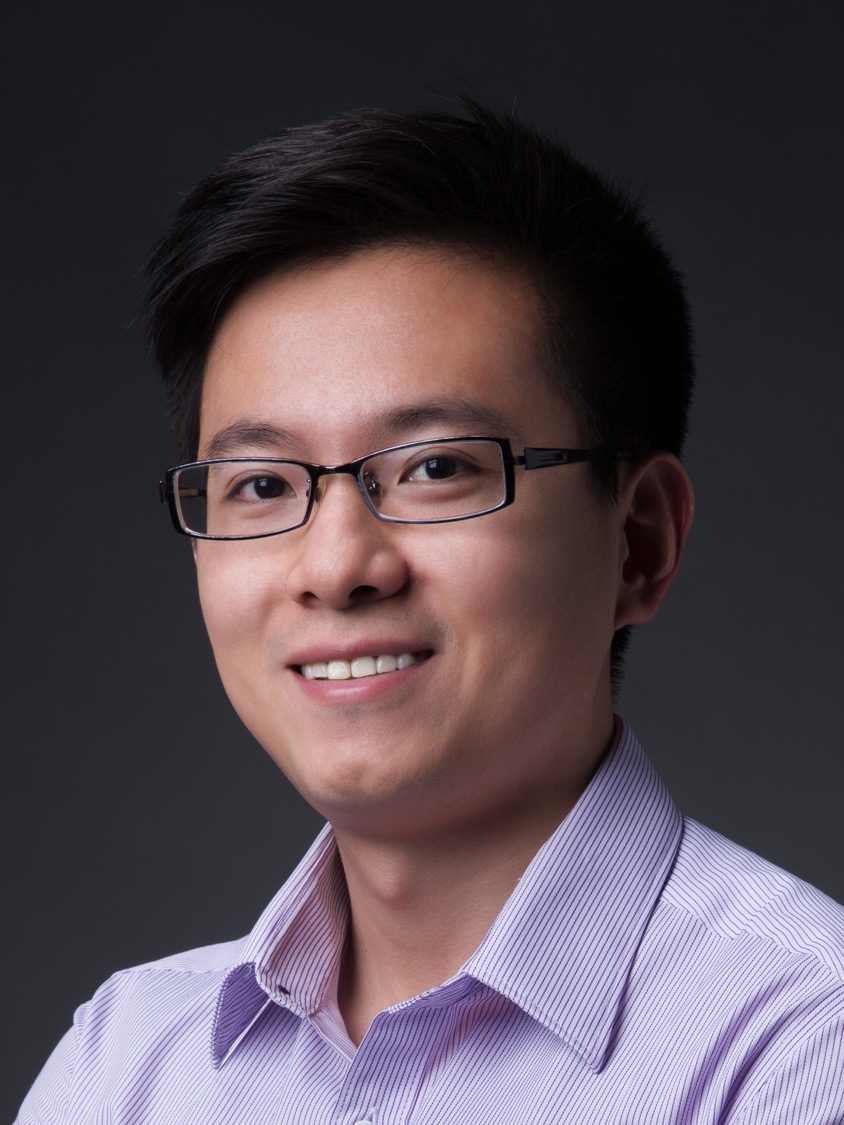}}]{Xiu-Shen Wei} is a Professor with the School of Computer Science and Engineering, Southeast University, China. He was a Program Chair for the workshops associated with ICCV, IJCAI, ACM Multimedia, etc. He has also served as an Area Chair or Senior Program Member at CVPR, AAAI, IJCAI, ICME, BMVC, a Guest Editor of Pattern Recognition Journal, and a Tutorial Chair for Asian Conference on Computer Vision (ACCV) 2022.
\end{IEEEbiography}

\begin{IEEEbiography}[{\includegraphics[width=1in,height=1.25in,clip,keepaspectratio]{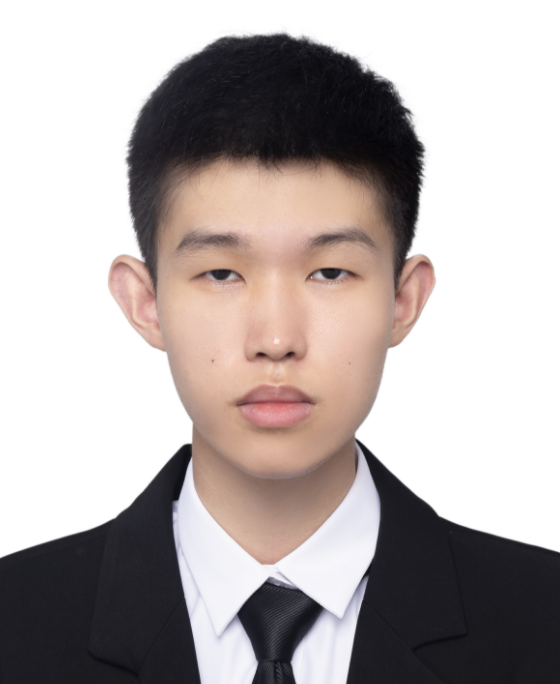}}]{Yang Shen} is a Ph.D Candidate under the supervision of Prof. Xiu-Shen Wei with School of Computer Science and Engineering, Nanjing University of Science and Technology, China. His research interests lie in deep learning and computer vision.
\end{IEEEbiography}

\begin{IEEEbiography}[{\includegraphics[width=1in,height=1.25in,clip,keepaspectratio]{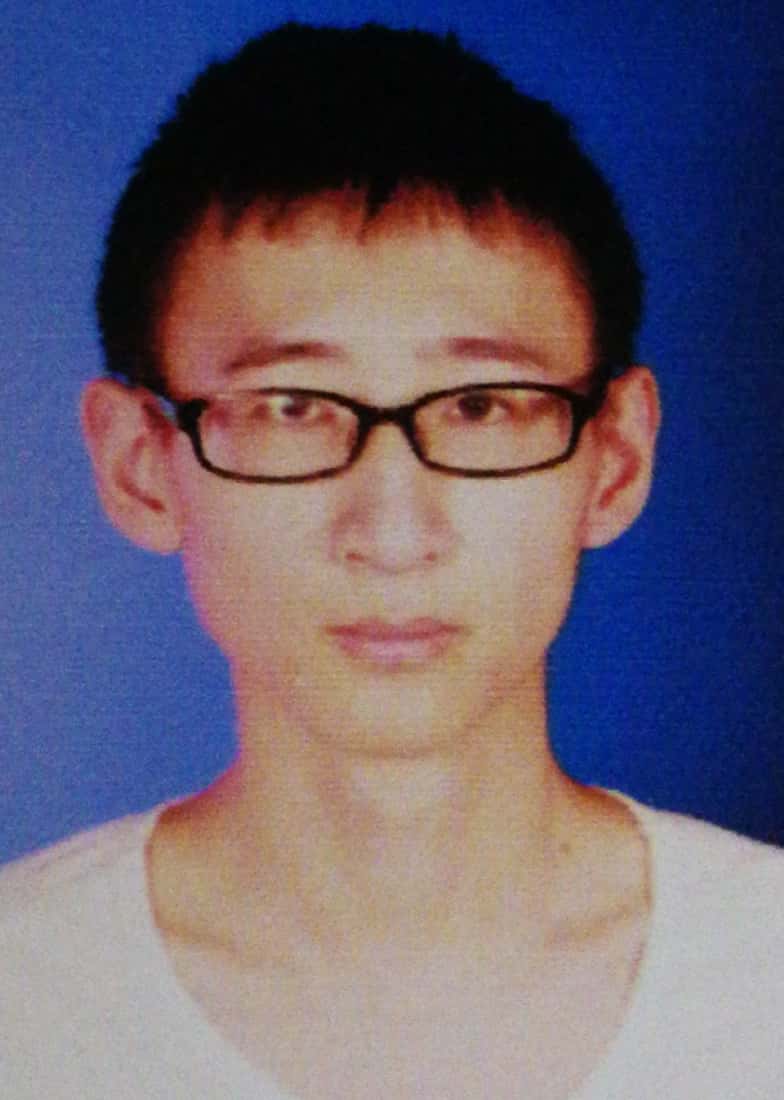}}]{Xuhao Sun} is a Master Student with School of Computer Science and Engineering, Nanjing University of Science and Technology, China. His main research interests are Computer Vision and Machine Learning.
\end{IEEEbiography}

\begin{IEEEbiography}[{\includegraphics[width=1in,height=1.25in,clip,keepaspectratio]{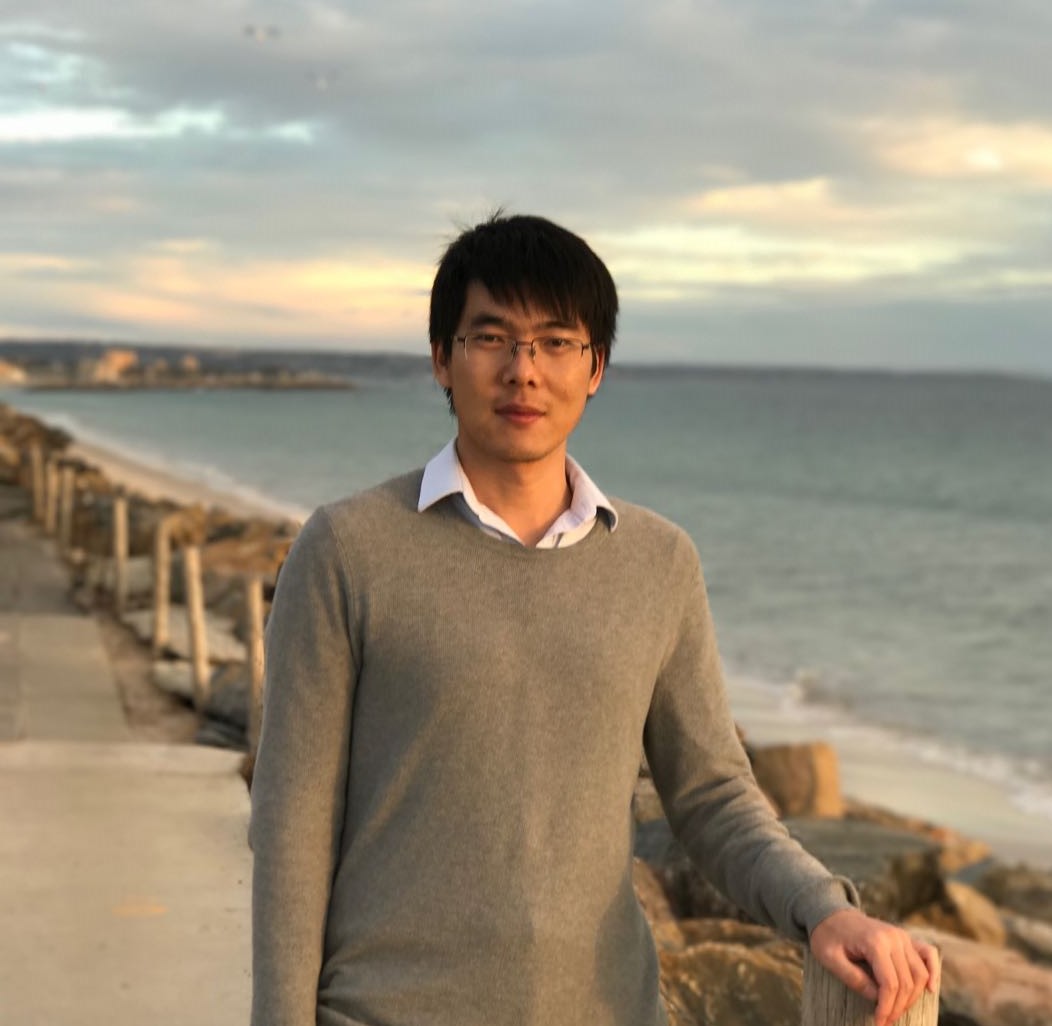}}]{Peng Wang} is a Professor with School of Computer Science and Engineering, University of Electronic Science and Technology of China. He holds a PhD from School of Information Technology and Electrical Engineering at The University of Queensland. Dr. Peng Wang's primary research focus centers around computer vision and machine learning, with a particular emphasis on label-efficient deep learning techniques.% His research interests encompass areas such as few-shot learning, long-tailed learning, and self-supervised learning.
\end{IEEEbiography}

\begin{IEEEbiography}[{\includegraphics[width=1in,height=1.25in,clip,keepaspectratio]{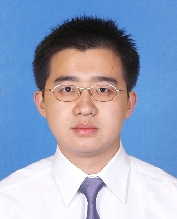}}]{Yuxin Peng} is currently the Boya Distinguished Professor with the with Wangxuan Institute of Computer Technology, and National Key Laboratory for Multimedia Information Processing, Peking University. He received the Ph.D. degree in computer applied technology from Peking University in 2003. He has authored over 180 papers, including more than 90 papers in the top-tier journals and conference proceedings. He has submitted 48 patent applications and been granted 39 of them. His current research interests mainly include cross-media analysis and reasoning, image and video recognition and understanding, and computer vision. He led his team to win the First Place in video instance search evaluation of TRECVID ten times in the recent years. He won the First Prize of the Beijing Technological Invention Award in 2016 (ranking first) and the First Prize of the Scientific and Technological Progress Award of Chinese Institute of Electronics in 2020 (ranking first). He was a recipient of the National Science Fund for Distinguished Young Scholars of China in 2019, and the best paper award at MMM 2019 and NCIG 2018. He serves as the associate editor of IEEE TMM, TCSVT, etc.
\end{IEEEbiography}

\end{document}


%\title{Large-Scale Fine-Grained Image Retrieval with Attribute-Aware Hash Codes}
%\title{Attribute-Aware Deep Hashing with Self-Consistency for Large-Scale Fine-Grained Image Retrieval (Appendices)}
\title{Appendices}

\author{
	Xiu-Shen Wei,~\IEEEmembership{Member,~IEEE}, Yang Shen, Xuhao Sun, Peng Wang,~\IEEEmembership{Member,~IEEE}, Yuxin Peng,~\IEEEmembership{Senior Member,~IEEE}
\IEEEcompsocitemizethanks{\IEEEcompsocthanksitem {X.-S. Wei is with School of Computer Science and Engineering, and Key Laboratory of New Generation Artificial Intelligence Technology and Its Interdisciplinary Applications, Southeast University, China. Y. Shen and X. Sun are with School of Computer Science and Engineering, Nanjing University of Science and Technology, China. P. Wang is with School of Computer Science and Engineering, University of Electronic Science and Technology of China, China. Y. Peng is with Wangxuan Institute of Computer Technology, and National Key Laboratory for Multimedia Information Processing, Peking University, China.}
}
%~\IEEEmembership{Member,~IEEE,}
}

\markboth{ACCEPTED BY IEEE TPAMI}%
{Wei \MakeLowercase{\textit{et al.}}: \textsc{A$^2$-Net}: Learning Attribute-Aware Hash Codes for Large-Scale Fine-Grained Image Retrieval}

\maketitle

%\tableofcontents

\IEEEdisplaynontitleabstractindextext

\IEEEpeerreviewmaketitle

\appendices
\section{Architecture of Deconvolutional Network}
We present the detailed architecture of the deconvolutional network $\Psi_{\rm Deconv}(\cdot;\Pi)$ in \textsc{A$^2$-Net$^{++}$}, cf. Section~3.4.2 of the paper, which is a basic and light-weight model as follows.
\begin{table}[h!]
\centering
\small
\caption{Detailed architecture specifications of $\Psi_{\rm Deconv}(\cdot;\Pi)$. The parameters $[d_{\rm in}, {\rm kern}\times {\rm kern}, d_{\rm out}]$ represent the values of input channels, deconvolution kernel size, and output channels.}
\vspace{-1em}
\begin{tabular}{c|c}
\toprule
Layer & Parameters            \\
\hline
1 & $[1024, 7\times7, 256]$ \\
2 & $[256, 4\times4, 128]$  \\
3 & $[128, 4\times4, 64]$   \\
4 & $[64, 4\times4, 32]$    \\
5 & $[32, 4\times4, 16]$    \\
6 & $[16, 4\times4, 3]$    \\
\bottomrule
\end{tabular}
\end{table}

\section{Additional experimental results of other comparison methods}

Apart from ExchNet, DSaH~\cite{jin2020deep} is another fine-grained hashing method which has achieved good retrieval accuracy. For fair comparisons, we strictly control empirical settings as the same as those of~\cite{jin2020deep} and compare the results of our models with its results and three following methods, \ie, DPSH~\cite{li2015feature}, DTQ~\cite{liu2018deep} and HBMP~\cite{cakir2018hashing}.

Specifically, we follow the settings of DSaH~\cite{jin2020deep} and conduct experiments on two fine-grained datasets, \ie, \textit{CUB200-2011}~\cite{WahCUB200_2011} and \textit{Stanford Dogs}~\cite{khosla2011novel}. In concretely, \textit{Stanford Dogs} consists of 20,580 images in 120 classes while each class contains about 150 images. The dataset is divided into the train set (100 images per class) and the test set (totally 8,580 images for all categories). \textit{CUB200-2011} contains 11,788 bird images from 200 bird species and is officially split into 5,994 images for training and 5,794 images for test. We use AlexNet as backbone and it is not fine-tuned on each dataset.

As shown in Table~\ref{table:results_appendix}, our proposed \textsc{A$^2$-Net} and {\textsc{A$^2$-Net$^{++}$}} models significantly outperform the other baseline methods on these two datasets by following the same settings of~\cite{jin2020deep}. In particular, compared with DSaH~\cite{jin2020deep}, our \textsc{A$^2$-Net} ({\textsc{A$^2$-Net$^{++}$}}) achieves 11.4\% (11.9\%) and 7.2\% (8.2\%) improvements on \textit{Stanford Dogs} and \textit{CUB200-2011} in average.

%Our source codes and pre-trained models are available at: \url{https://anonymous.4open.science/r/A-2-Net-9460}

%=======================table9=============================
\begin{table}[t!]
	\centering
	\small
	\caption{Additional comparisons of retrieval accuracy (\% mAP) with DPSH~\cite{li2015feature} on two benchmark fine-grained datasets. Best results are marked in bold.}
	\vspace{-1em}
	\setlength{\tabcolsep}{0.35pt}
	\begin{tabular}{c|c|cccc|cc}
\toprule
Datasets & \multicolumn{1}{c|}{\# bits} & \multicolumn{1}{c}{DPSH} & \multicolumn{1}{c}{DTQ} & \multicolumn{1}{c}{HBMP} & \multicolumn{1}{c|}{DSaH} & \multicolumn{1}{c}{\textbf{\textsc{A$^2$-Net}}} & \multicolumn{1}{c}{\textbf{\textsc{A$^2$-Net$^{++}$}}} \\
\hline
\multirow{4}[2]{*}{\textit{CUB200-2011}} & 12 & 17.7 & 18.5 & 19.0  & 24.4 & 36.6 & \textbf{37.2} \bigstrut[t]\\
& 24 & 22.1 & 18.7 & 23.8 & 28.7 & 44.8 & \textbf{45.2}\\
& 32 & 26.5 & 18.7 & 28.7 & 36.3 & 46.9 & \textbf{47.3}\\
& 48 & 31.5 & 18.8 & 32.8 & 40.8 & 47.4 & \textbf{48.1}\\
\hline
\multirow{4}[2]{*}{\textit{Stanford Dogs}} & 12 & ~~7.2 & ~~7.3 & ~~8.9 & 14.2 & 19.2 & \textbf{20.1} \bigstrut[t]\\
& 24 & ~~7.6 & 11.3 & 10.9 & 20.9 & 27.2 & \textbf{28.3}\\
& 32 & ~~8.4 & 15.4 & 14.2 & 23.2 & 32.5 & \textbf{32.9}\\
& 48 &  ~~7.9 & 18.3 & 16.8 & 28.5 & 36.7 & \textbf{38.6}\\	
\bottomrule
\end{tabular}%
	\label{table:results_appendix}%
\end{table}%

%=======================table10=============================
\begin{table}[t!]
	\centering
	\small
	\caption{Additional comparisons of retrieval accuracy (\% mAP) with FISH~\cite{tipxinshunTIP} on three benchmark fine-grained datasets. Best results are marked in bold.}
	\vspace{-1em}
	\begin{tabular}{c|c|c|cc}
\toprule
Datasets                              & \# bits & \multicolumn{1}{c|}{FISH} & \multicolumn{1}{c}{\textbf{\textsc{A$^2$-Net}}} & \multicolumn{1}{c}{\textbf{\textsc{A$^2$-Net$^{++}$}}} \\
\hline
\multirow{4}{*}{\textit{CUB200-2011}} & 12      & 76.77                    & 77.16                                           & \textbf{77.96}   \bigstrut[t]                                      \\
                                      & 24      & 79.93                    & 80.69                                           & \textbf{81.25}                                         \\
                                      & 32      & 80.09                    & 81.04                                           & \textbf{81.66}                                         \\
                                      & 48      & 80.88                    & 81.73                                           & \textbf{82.10}                                         \\
\hline
\multirow{4}{*}{\textit{Aircraft}}    & 12      & 88.29                    & 88.93                                           & \textbf{89.50}  \bigstrut[t]                                       \\
                                      & 24      & 89.20                    & 89.54                                           & \textbf{89.92}                                         \\
                                      & 32      & 89.28                    & 89.62                                           & \textbf{89.98}                                         \\
                                      & 48      & 89.49                    & 89.81                                           & \textbf{90.14}                                         \\
\hline
\multirow{4}{*}{\textit{VegFru}}      & 12      & 79.17                    & 79.53                                           & \textbf{80.87}      \bigstrut[t]                                   \\
                                      & 24      & 85.33                    & 85.67                                           & \textbf{86.73}                                         \\
                                      & 32      & 85.43                    & 85.83                                           & \textbf{86.92}                                         \\
                                      & 48      & 85.51                    & 86.41                                           & \textbf{87.11}                                        
\\	
\bottomrule
\end{tabular}%
	\label{table:results_appendix2}%
\end{table}%

Moreover, we also follow the entire empirical settings of FISH~\cite{tipxinshunTIP} and conduct experiments on \textit{CUB200-2011}~\cite{WahCUB200_2011}, \textit{Aircraft}~\cite{airplanes}, and \textit{VegFru}~\cite{vegfru}. We use ResNet-50 as backbone for fair comparisons. As results reported in Table~\ref{table:results_appendix2}, it can be consistently observed that our significant improvements of \textsc{A$^2$-Net} and \textsc{A$^2$-Net$^{++}$} over FISH on varied lengths of hash bits on these three fine-grained datasets. Especially when the average retrieval accuracy is as high as 85\% on these datasets, our models (\eg, \textsc{A$^2$-Net$^{++}$}) can still achieve an accuracy improvement of more than 1.2\% in average.

\iffalse
\ifCLASSOPTIONcompsoc
  \section*{Acknowledgments}
\else
  \section*{Acknowledgment}
\fi

The authors would like to thank...

\ifCLASSOPTIONcaptionsoff
  \newpage
\fi

\fi

%\section*{Acknowledgments}

%The authors would like to thank the editor and the anonymous reviewers for their constructive comments. 
%This work was supported by National Key R\&D Program of China (2021YFA1001100), National Natural Science Foundation of China under Grant (62272231), Natural Science Foundation of Jiangsu Province of China under Grant (BK20210340), the Fundamental Research Funds for the Central Universities (No. NJ2022028), and CAAI-Huawei MindSpore Open Fund. We gratefully acknowledge the support of MindSpore, CANN (Compute Architecture for Neural Networks) and Ascend AI Processor used for this research.

\bibliographystyle{IEEEtran}
\bibliography{FGIA_survey_full_A2Net}
